\documentclass{article}

\usepackage{arxiv}

\usepackage[utf8]{inputenc} % allow utf-8 input
\usepackage[T1]{fontenc}    % use 8-bit T1 fonts
\usepackage{hyperref}       % hyperlinks
\usepackage{url}            % simple URL typesetting
\usepackage{booktabs}       % professional-quality tables
\usepackage{amsfonts}       % blackboard math symbols
\usepackage{nicefrac}       % compact symbols for 1/2, etc.
\usepackage{microtype}      % microtypography
\usepackage{lipsum}
\usepackage{graphicx}
\usepackage{makecell}

\graphicspath{ {./images} }

\usepackage[caption=false,font=normalsize,labelfont=sf,textfont=sf]{subfig}
\usepackage{amsmath}
\usepackage{siunitx}
\usepackage[numbers]{natbib}

\title{Theory-guided Pseudo-spectral Full Waveform Inversion via Deep Neural Networks}

\author{
  Christopher Zerafa\\
  Department of Geosciences\\
  University of Malta\\
  Msida, Malta\\
  \texttt{christopher.zerafa.08@um.edu.mt} \\
   \And
   Pauline Galea \\
  Department of Geosciences\\
  University of Malta\\
  Msida, Malta\\
  \texttt{pauline.galea@um.edu.mt} \\
   \AND
   Cristiana Sebu \\
  Department of Geosciences\\
  University of Malta\\
  Msida, Malta\\
   \texttt{cristiana.sebu@um.edu.mt} 
}

\begin{document}
\maketitle
\begin{abstract}
  Full-Waveform Inversion seeks to achieve a high-resolution model of the subsurface through the application of multi-variate optimization to the seismic inverse problem. Although now a mature technology, FWI has limitations related to the choice of the appropriate solver for the forward problem in challenging environments requiring complex assumptions, and very wide angle and multi-azimuth data necessary for full reconstruction are often not available.
    
  Deep Learning techniques have emerged as excellent optimization frameworks. Data-driven methods do not impose a wave propagation model and are not exposed to modelling errors. On the contrary, deterministic models are governed by the laws of physics. 
  
  Seismic FWI has recently started to be investigated as a Deep Learning framework. Focus has been on the time-domain, while the pseudo-spectral domain has not been yet explored. However, classical FWI experienced major breakthroughs when pseudo-spectral approaches were employed. This work addresses the lacuna that exists in incorporating the pseudo-spectral approach within Deep Learning. This has been done by re-formulating the pseudo-spectral FWI problem as a Deep Learning algorithm for a theory-driven pseudo-spectral approach. A novel Recurrent Neural Network framework is proposed. This is qualitatively assessed on synthetic data, applied to a two-dimensional Marmousi dataset and evaluated against deterministic and time-based approaches.
  
  Pseudo-spectral theory-guided FWI using RNN was shown to be more accurate than classical FWI with only 0.05 error tolerance and 1.45\% relative percent-age error. Indeed, this provides more stable convergence, able to identify faults better and has more low frequency content than classical FWI. Moreover, RNN was more suited than classical FWI at edge detection in the shallow and deep sections due to cleaner receiver residuals.

\end{abstract}

\section{Introduction}
Full waveform inversion (FWI) endeavours to attain a high-resolution representation of subsurface structures through the application of multivariate optimization to the seismic inverse problem \cite{Virieux2009}. The foundational role of optimization theory within FWI stems from its essence in reconstructing the parameters of the investigated system from indirect observations, which are bound by a forward modelling process \cite{Tarantola2005}. It is worth noting that the selection of the forward problem holds a consequential sway on the precision of the FWI outcome. Particularly intricate environments necessitate more intricate assumptions to elucidate the physical connection between data and observations. However, this complexity does not necessarily guarantee heightened levels of accuracy \cite{Morgan2013}. Additionally, the data employed to construct the mapping of measurements for ground-truth representation often falls short of being optimal. Comprehensive solutions demand the integration of wide-angle and multi-azimuth data to enable complete reconstruction of the inverse problem \cite{Morgan2013}, yet this requisite is rarely met.

In recent times, the domain of deep learning has emerged as a potent paradigm for tackling inverse problems \cite{Elshafiey1991,Adler2017a}. These deep learning-based processes for waveform inversion exist at the crossroads of data-driven and theory-guided methodologies \cite{Sun2019}. Unlike theory-guided methods, data-driven techniques do not impose a wave propagation model as a priori knowledge. Neural network weights are entirely trainable, necessitating ample training datasets for effective inversion \cite{Sun2019a}. However, due to their substantial degree of flexibility, they are less susceptible to modelling errors compared to traditional FWI algorithms \cite{Wu2018}.

While the application of deep learning techniques to FWI is already in use, the focus has predominantly been on the time-domain approach \cite{Sun2019a}. Nevertheless, it is important to acknowledge that classical FWI witnessed transformative breakthroughs through pseudo-spectral strategies \cite{Sirgue2006}, enabling the method to transcend academic boundaries and be deployed with tangible success on authentic datasets \cite{Sirgue2009}. In this paper, we investigate whether we can extend research within pseudo-spectral FWI and derive the inversion as a theory-guided Deep Neural Network (DNN). To current knowledge, there is no prior work investigating the pseudo-spectral inversion within a theory-guided DNN framework. Thus, our primary goal is to explore and develop a novel approach known as theory-guided pseudo-spectral FWI. We compare it with traditional approaches, examining its advantages and limitations. To accomplish this, we follow these steps:
\begin{enumerate}
    \item Re-casting FWI within a theory-derived inversion-based DNN. This is derived theoretically and assessed on synthetic data. 
    \item Validating results against classical deterministic FWI.
    \item Analysing the limitations of the approach and discussing future potential developments.
\end{enumerate}

The paper will proceed as follows, in Section 2 we provide the mathematical foundation for FWI and in Section 3 we introduce deep learning and how it can be recast as a Theory-guided Neural Network. In Section 4 we these our new framework on synthetic data and compare against classical FWI. In the final Section we discuss the potential for this work, its limitations and future consideration and provide a conclusion.

\section{Theoretical Background}
The forward problem in FWI is based on the wave equation. It is a second order, partial differential equation involving both time and space derivatives. For an isotropic medium is given by:
\begin{equation}
\frac{1}{c(\mathbf{m})^2} \frac{\partial^2p(\mathbf{m},t)}{\partial t^2} - \nabla^2p(\mathbf{m},t) = s(\mathbf{m},t),
\end{equation}
where $p(\mathbf{m},t)$ is the pressure wave-field, $c(\mathbf{m})$ is the acoustic $p$-wave velocity and $s(\mathbf{m},t)$ is the source \cite{Igel2016}. To solve the wave equation numerically, it can be expressed as a linear operator. 
% Although the data $\mathbf{d}$ and model $\mathbf{m}$ are not linearly related, the wave-field $p(\mathbf{m},t)$ and the sources $s(\mathbf{m},t)$ are linearly related by the equation:
% \begin{equation}
% \mathbf{A}p(\mathbf{m},t) = s(\mathbf{m},t),
% \end{equation}
% where $p(\mathbf{m},t)$ is the pressure wave-field produced by a source $s(\mathbf{m})$ and $\mathbf{A}$ is the numerical implementation of the operator:
% \begin{equation}
% \frac{1}{c(\mathbf{m})^2} \frac{\partial^2}{\partial t^2} - \nabla^2,
% \end{equation}

Based on the Born approximation in scattering theory \cite{Born1980}, consider the first model calculated to be $\boldsymbol{x}_0$. After the first pass via forward modelling, the model needs to be updated by the model parameter perturbation $\Delta \boldsymbol{x}_0$. This newly updated model is then used to calculate the next update and the procedure continues iteratively until the computed mode is close enough to the true model based on a residual threshold criterion. At each iteration $k$, the misfit function $\phi(\boldsymbol{x}_k )$ is calculated from model $\boldsymbol{x}_{k-1}$ of the previous iteration giving:
\begin{equation}\label{eq:misfit_next_step}
	\phi(\boldsymbol{x}_k)=\phi(\boldsymbol{x}_{k-1} + \Delta\boldsymbol{x}_k).
\end{equation}
Assuming that the model perturbation is small enough with respect to the model, Equation~\ref{eq:misfit_next_step} can be expanded via Taylor expansions up to second orders as:
\begin{equation}\label{eq:misfit_taylor_expansion}
	\phi(\boldsymbol{x}_k) = \phi(\boldsymbol{x}_{k-1}) + \delta\boldsymbol{x}^T\frac{\partial\phi}{\partial x}+\frac{1}{2}\delta\boldsymbol{x}^T\frac{\partial^2\phi}{\partial x^2}\delta\boldsymbol{x}.
\end{equation}
Taking the derivative of Equation~\ref{eq:misfit_taylor_expansion} and minimizing to determine the model update leads to:
\begin{equation}\label{eq:hessian}
	\partial\boldsymbol{x}\approx-\boldsymbol{H}^{-1}\nabla_x\phi,
\end{equation}
where $\boldsymbol{H}=\frac{\partial^2\phi}{\partial x^2}$ is the Hessian matrix and $\nabla_x\phi$ the gradient of misfit function evaluated at $\boldsymbol{x}_0$. The Hessian matrix is symmetric and represents the curvature trend of the misfit function.

A common technique employed within the forward modelling stage is to perform modelling in pseudo-spectral domain rather than the time domain. The most common domain is the Fourier domain \cite{Igel2016} and implementation is generally achieved via the Fast Fourier Transform (FFT) \cite{cooley1965algorithm}. 

% \begin{figure*}[b]
% 	\centering
% 	\includegraphics[width=0.95\linewidth]{fwi_workflow.png}
% 	\caption{Schematic of a FWI workflow solved as an iterative optimisation process.}
% 	\label{fig:algo_schem_fwi}
% \end{figure*}

After forward modelling the data in the pseudo-spectral domain, the objective is to seek to minimize the difference between the observed data and the modelled data. The misfit between the two datasets is known as the objective- or cost-function $\emph{J}$. The most common cost function is given by the $L_2$-norm of the data residuals:
% \begin{equation}
% \emph{J}(\mathbf{m}) = \frac{1}{2}{\left|| d - F(\mathbf{m}) \right||}^2_D,
% \end{equation}
\begin{equation}
    \emph{J}(\mathbf{m}) = \frac{1}{2}\left[ {\left|| \mathbf{d} - F(\mathbf{m}) \right||}^2_D + \lambda\left||\mathbf{m}\right||^2_M \right] ,
\end{equation}
where $D$ indicates the data domain given by $n_s$ sources and $n_r$ receivers, $M$ is the model domain, and $\lambda$ is a regularization parameter. The misfit function $\emph{J}$ can be minimized with respect to the model parameters $m$ if the gradient is zero, namely:
\begin{equation}
\nabla\emph{J} = \frac{\partial\emph{J}}{\partial \mathbf{m}} = 0,
\end{equation}

% Minimising the misfit function is generally achieved via a linearised iterative optimisation scheme based on the Born approximation in scattering theory \cite{Born1980}. 
% The inversion algorithm starts with an initial estimate of the model $\mathbf{m}_0$. After the first pass via forward modelling, the model is updated by the model parameter perturbation $\Delta \mathbf{m}_0$. This newly updated model is then used to calculate the next update and the procedure continues iteratively until the computed model is close enough to the observations based on a residual threshold criterion.
At each iteration $k$, assuming small enough model perturbation and using Taylor Expansion up to second orders, the misfit function $\emph{J}(\mathbf{m}_k)$ is calculated from the previous iteration model $\mathbf{m}_{k-1}$ as:
%  by:
% \begin{equation} \label{eq:misfit_k-1}
% \emph{J}(\mathbf{m}_k) = \emph{J}(\mathbf{m}_{k-1} + \Delta \mathbf{m}_k),
% \end{equation}
\begin{equation} \label{eq:tay_exp}
    \emph{J}(\mathbf{m}_k) = \emph{J}(\mathbf{m}_{k-1}) 
		+ \delta \mathbf{m}^T_{k-1}
		\frac{\partial\emph{J}}{\partial\mathbf{m}_{k-1}}
		+ \frac{1}{2}
		\delta\mathbf{m}^{2T}_{k-1}
		\frac{\partial^2\emph{J}}{\partial\mathbf{m}^2_{k-1}},
\end{equation}

\section{Proposed FWI as a Theory-Guided DNN}
Neural networks are a subset of tools in artificial intelligence which when applied to inverse problems can approximate the non-linear function of the inverse problem $F^{-1}:D\rightarrow M$. That is, using a neural network, a non-linear mapping can be learned to minimize:
\begin{equation}
\left||\mathbf{m} - g_{\theta}(\mathbf{d})\right||^2 ,
\end{equation}
where $\theta$ the large data set of pairs $(\mathbf{m}, \mathbf{d})$ used for the learning process \cite{Adler2017a}. Recurrent Neural Networks (RNNs) are a class of neural networks designed to process sequences of data by maintaining an internal memory to capture temporal dependencies. They are particularly adept at handling tasks like natural language processing and time series analysis \cite{Olah2015}.  

% Classic feed-forward NNs, or Multi-Layer Perceptrons (MLP), do not allow for cyclical/recurrent connections between neurons. If this condition is relaxed, we obtain Recurrent Neural Networks. This difference is illustrated in Figure~\ref{fig:MLP_vs_RNN}. Although this might seem trivial, recurrent connections allow a ``memory'' of previous inputs to persist in the network’s state \cite{Graves2012}. Through results of the Universal Approximation Theorem, \cite{Hammer2000} prove that a 
% ``RNN with a sufficient number of hidden units can approximate any measurable sequence-to-sequence mapping to arbitrary accuracy''.

% \begin{figure}[ht!]
% 	\centering
% 	\includegraphics[width=0.8\linewidth]{MLP_vs_RNN.png}
% 	\caption[Difference between NNs and RNN architectures.]{Difference between feed-forward NNs and RNN architectures are Red and Blue cyclical connections. Red are intra-layer connections, whereas Blue are across layers.}
% 	\label{fig:MLP_vs_RNN}
% \end{figure}

\subsection[Long Short-Term Memory]{Long Short-Term Memory (LSTM)}
Standard RNN architectures suffer from the vanishing gradient problem \cite{Hochreiter1991}. Namely, the sensitivity of deeper neurons either decays or blows up exponentially as it passes through the recurrent connections \cite{Graves2012}. 

\begin{figure}[!ht]
	\centering
	\subfloat[The repeating module in a standard RNN contains a single layer.\label{fig:LSTM_module_RNN}]{\includegraphics[width=0.8\linewidth]{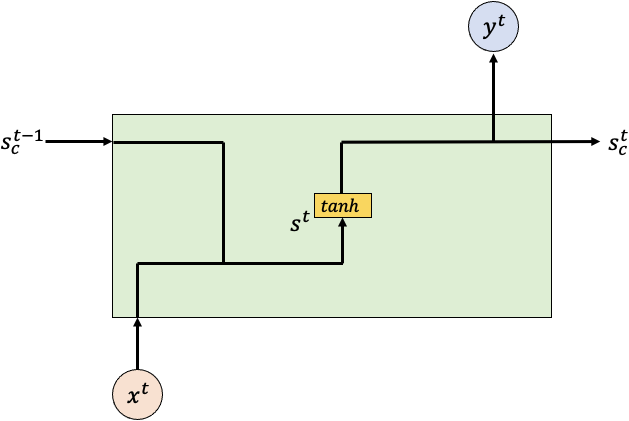}}
 \\
	\subfloat[A LSTM memory block has an additional three gates – Input, Output and Forget Gate (red) and a cell state block (blue).\label{fig:LSTM_module_LSTM}]{\includegraphics[width=0.8\linewidth]{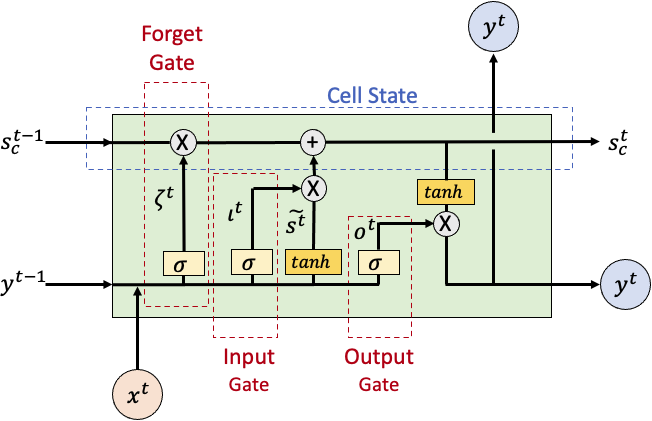}}
	\caption[Comparison between RNN and LSTM blocks.]{Comparison between RNN and LSTM blocks. Adapted from \cite{Olah2015}.}        
	\label{fig:RNN_LSTM_blocks}
\end{figure}

In 1997, \cite{Hochreiter1997} introduce a modified architecture type known as Long Short-Term Memory (LSTM) which mitigates the vanishing gradient problem. This NN introduces a set of recurrent connections known as memory blocks (the input, output and forget gates) and a cell state. Figure~\ref{fig:LSTM_module_RNN} shows a standard RNN with a single $tanh$ layer. Figure~\ref{fig:LSTM_module_LSTM} shows the LSTM chain structure but with the additional four interaction layers. Mathematical detail for each of these components is given in Appendix~\ref{sec:app_theory_Individual_RNN_Components}.

\subsection{LSTM as a Substitute for Wave Propagation}
\begin{figure}[ht!]
	\centering
	\subfloat[Unrolled form of acyclic graph of LSTM for FWI.\label{fig:LSTM_unrolled}]{\includegraphics[width=0.8\linewidth]{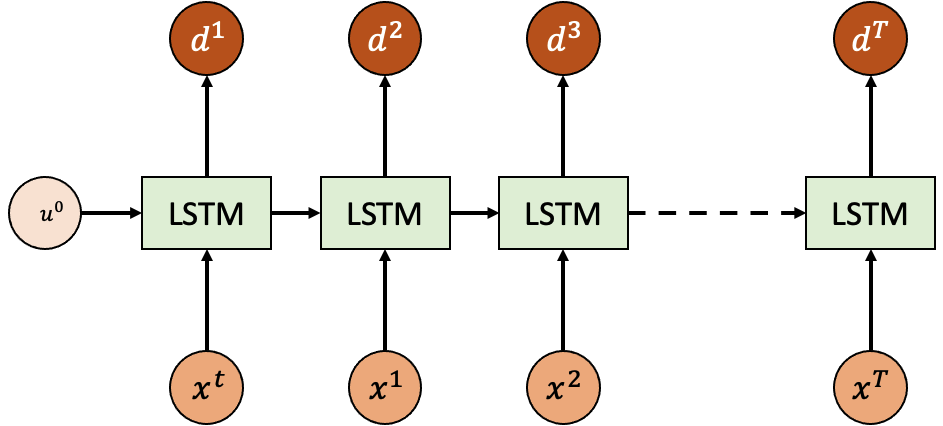}}
 \\
	\subfloat[Modified LSTM cell block supporting of forward modelling.\label{fig:LSTM_FWI_components}]{\includegraphics[width=0.8\linewidth]{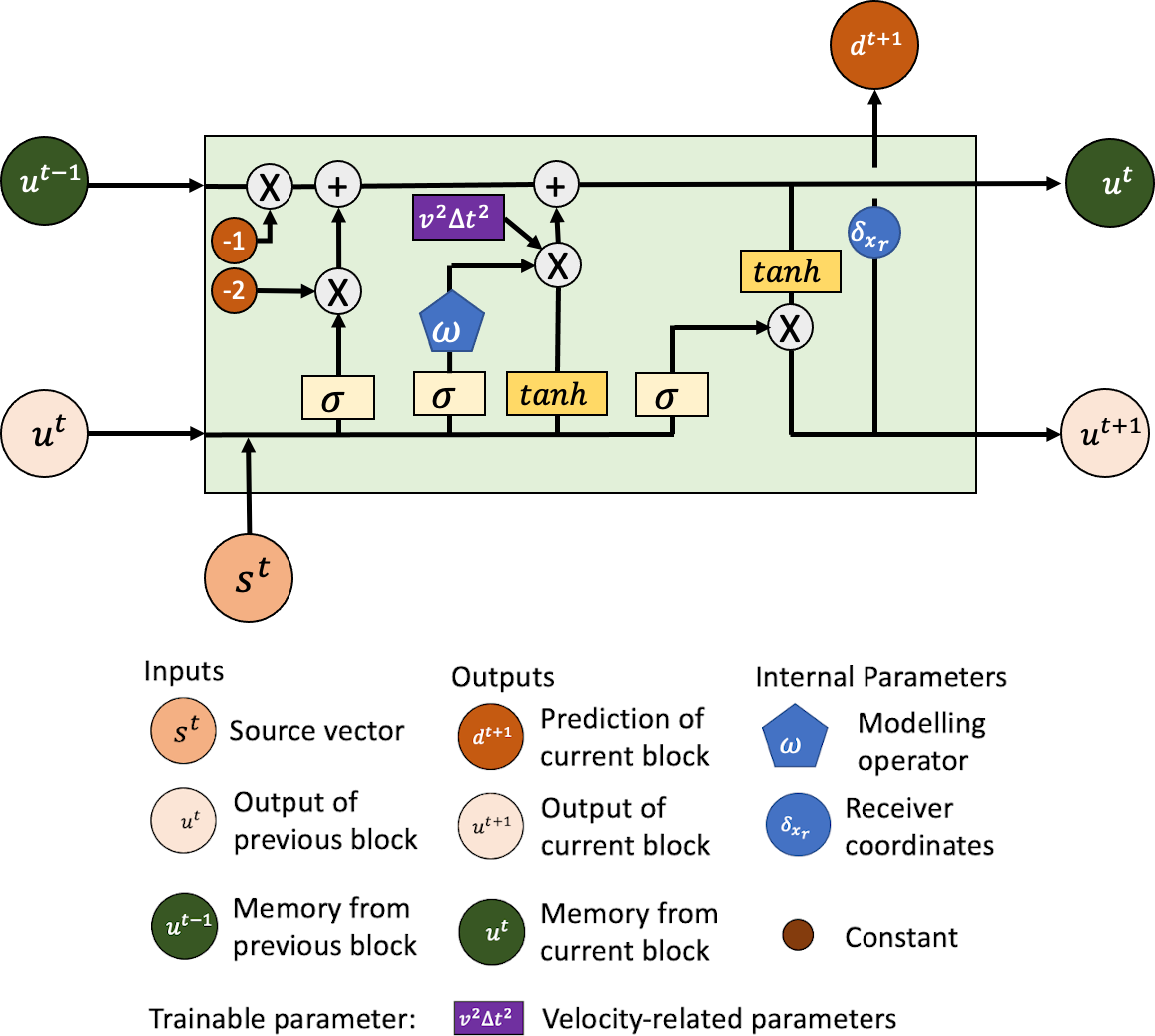}}
	\caption[Recasting of forward modelling within an LSTM]{Recasting of forward modelling of FWI within an LSTM deep learning framework. Adapted from \cite{Sun2019}.}        
	\label{fig:LSTM_FWI}
\end{figure}

Consider the discretized finite-difference stencil for wave propagation given as,
\begin{equation}
	p_j^{n+1}=\partial t^2 \left[c_j^2 \mathcal{F}^{-1} \left[k^2 \mathcal{P}_{\nu}^n\right]+s_j^n\right]+2p_j^n-p_j^{n-1},
\end{equation}
it is clear how pressure waves $p$ and source impulse $s$ at current time step $n$ are not affected by the future values $n+1$, but only dependent on the previous state of pressure at $n-1$. This is, by definition, a finite impulse with directed acyclic graph under graph theory definitions \cite{Thulasiraman2011}. With slight modification to the LSTM blueprint in Figure~\ref{fig:LSTM_module_LSTM}, a Deep Learning architecture supporting forward modelling can be cast as a LSTM cell that considers the pressure wave at time $n-1$, produces the modelled shot record at current time $n$ and stores this in memory for the next step $n+1$. Measuring all outputs at each moment in time would equal to the measurements of the wavefield locally at a geophone. This LSTM architecture is shown in Figure~\ref{fig:LSTM_unrolled} as an unrolled graph and Figure~\ref{fig:LSTM_FWI_components} as the building block components within an LSTM. 

The inputs to the LSTM cell are the source term at current time $s^t$, the wavefield at current $u_t$ and the previous time step $u_{t-1}$ stored in the memory of the LSTM. These wavefields are combined together with untrainable modelling operator $\omega$ and constants $-1$ and $-2$ to replicate the incremental time stepping in forward modelling. Deciding to model in time is equivalent to setting $\omega$ to the Laplacian, whereas setting it to calculate pseudo-spectral second-order derivates will lead to pseudo-spectral wavefield modelling. The trainable velocity-related parameter $v^2 \Delta t^2$ is applied to get the current modelled wavefield $u^{t+1}$. This is stored in memory, passed to the forget gate and receiver location discretization $\delta_{x_{r}}$ is applied to get the predicted outputs $d^{t+1}$. To train the velocity parameters, seismic shot records are provided as labelled data for training.

\section{Experimental Results}\label{sec:results_discussion}

\subsection{Forward Modelling using RNNs}\label{sec:results_forward_modelling_using_rnnd}
\begin{figure}[ht!]
    \centering
    \subfloat[Constant velocity model.\label{fig:rnn_2d_multisource_vel}]{%
    \includegraphics[width=0.3\linewidth]{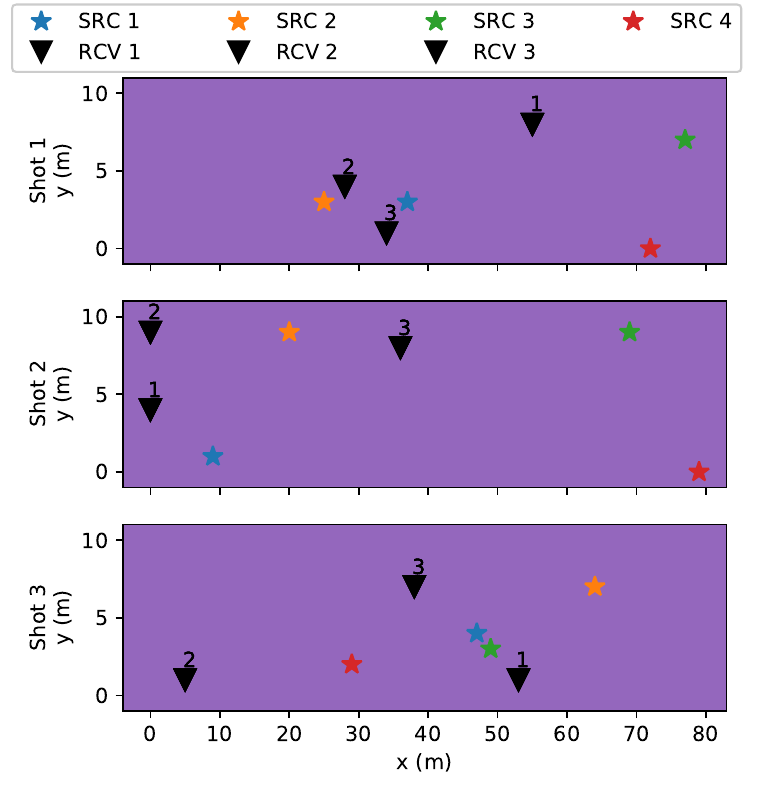}}
    \subfloat[Direct wave RNN forward modelling.\label{fig:rnn_2d_multisource_trace}]{%
    \includegraphics[width=0.5\linewidth]{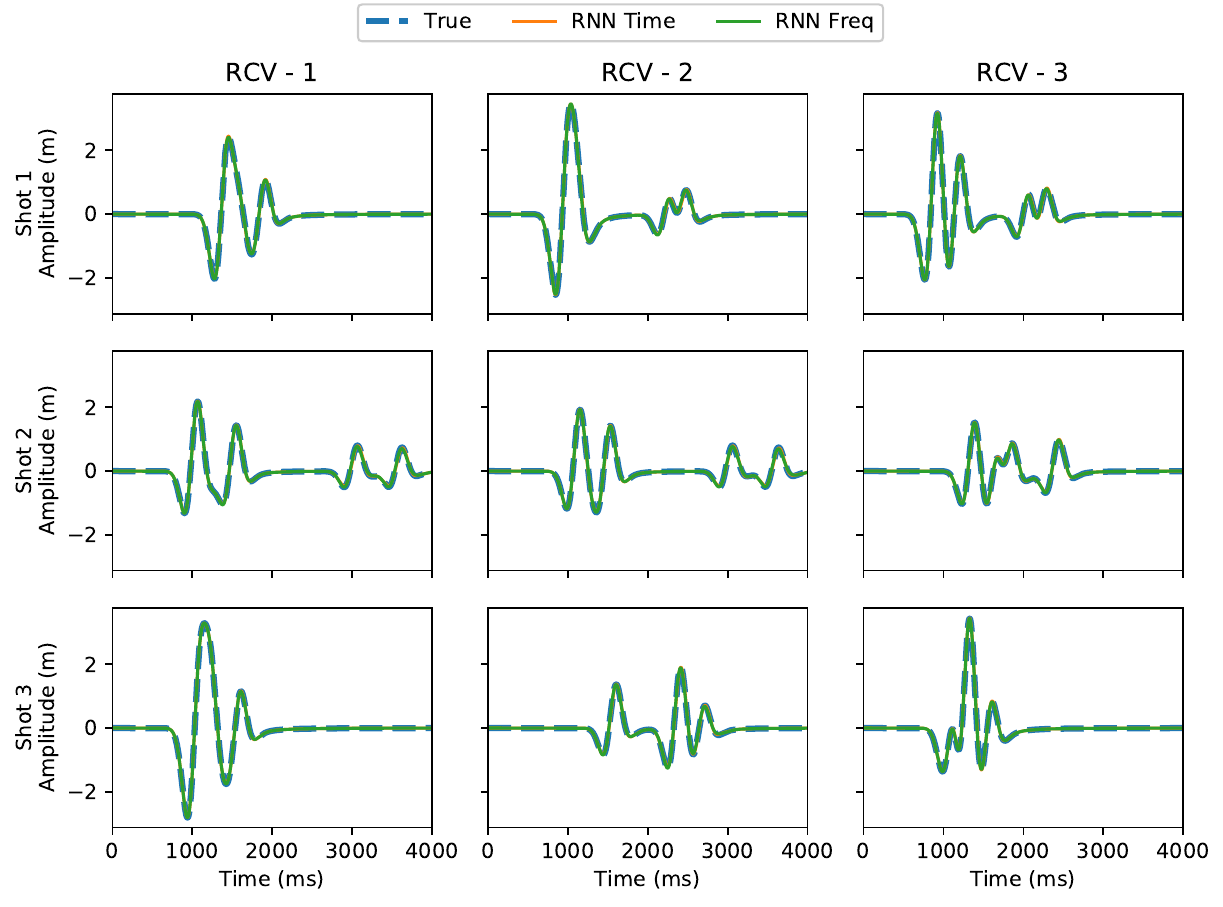}}
    \caption[Direct wave forward modelling for multi-source, multi-receiver geometry.]{Direct wave forward modelling for multi-source, multi-receiver geometry.}        
    \label{fig:rnn_2d_direct_multi}
\end{figure}

RNN should be able to model the different wave field components if it is to replace the forward modelling component. A 25Hz Ricker wavelet was propagated through a 2D 1500\si{ms^{-1}} constant velocity model (Figure~\ref{fig:rnn_2d_multisource_vel}) with a multi-source multi-receiver geometry setup. The 25Hz source wavelet goes into the hyper-resolution realm for FWI and is beyond the resolution that will be investigated on the synthetic model, however, this allows for gauging the limit of accuracy. This model setup was forward propagated for 5333 time-steps at 1ms, with a 10m grid spacing. Namely, this implies that 5333 LSTM cells were employed for the forward modelling. The resulting direct waves are illustrated in Figure~\ref{fig:rnn_2d_multisource_trace}, with True being the analytical solution calculated using a 2D Green’s function, RNN Time and RNN Freq are the RNN implementations for forward modelling using Time and Fourier spatial derivatives respectively. Qualitatively, there is no visible difference between either approach. 

\begin{figure}[ht!]
	\centering
	\subfloat[Step velocity model.\label{fig:rnn_2d_reflection_vel}]{\includegraphics[width=0.35\linewidth]{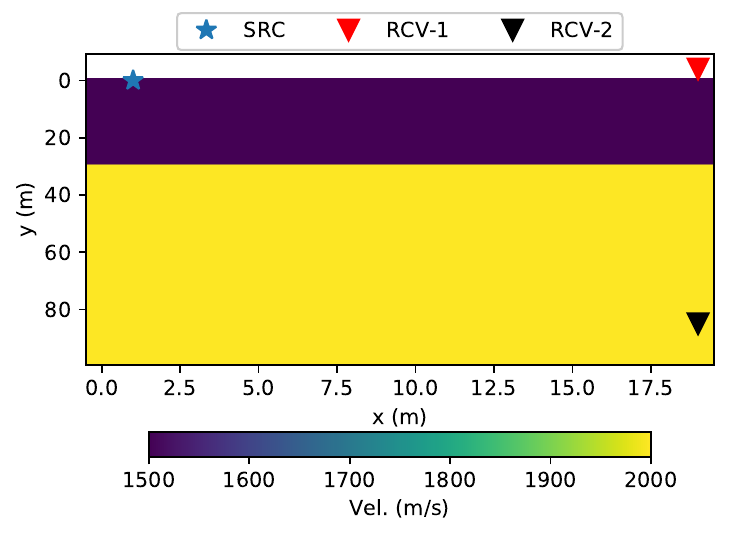}}
    \subfloat[Top: RCV-1 located at ground level reacts to direct arrival at 125ms and reflected arrival at 250ms. \newline Bottom: RCV-2 shows transmitted arrival.\label{fig:rnn_2d_reflection_wave}]{\includegraphics[width=0.5\linewidth]{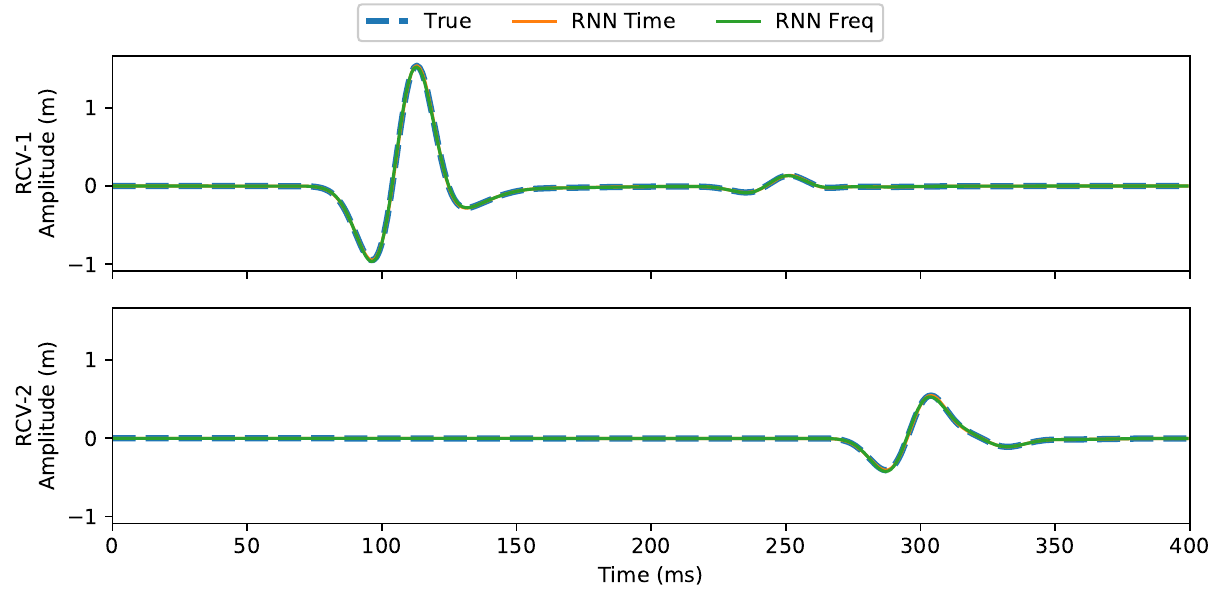}}
	\caption[Reflected and transmitted wave RNN forward modelling.]{Reflected and transmitted wave RNN forward modelling.}        
	\label{fig:rnn_2d_reflection}
\end{figure}

Reflected and transmitted arrivals were tested using a simple step velocity model ranging from 1500\si{ms^{-1}} to 2000\si{ms^{-1}} as shown in Figure~\ref{fig:rnn_2d_reflection_vel}. Figure~\ref{fig:rnn_2d_reflection_wave} is the forward-modelled wavefield for the two receiver locations (RCV-1 and RCV-2), top and bottom respectively. RCV-1 at ground level interacts with the direct wave at 125ms and the reflected wave at 250ms. RCV-2 is below the acoustic impedance layer at 30m and shows the transmitted wave. Comparing these to the analytical solution, either are able to model the wave components perfectly.

\begin{figure}[ht!]
\centering
\subfloat[Point-scattering velocity model.\label{fig:rnn_2d_scattered_vel}]{\includegraphics[width=0.35\linewidth]{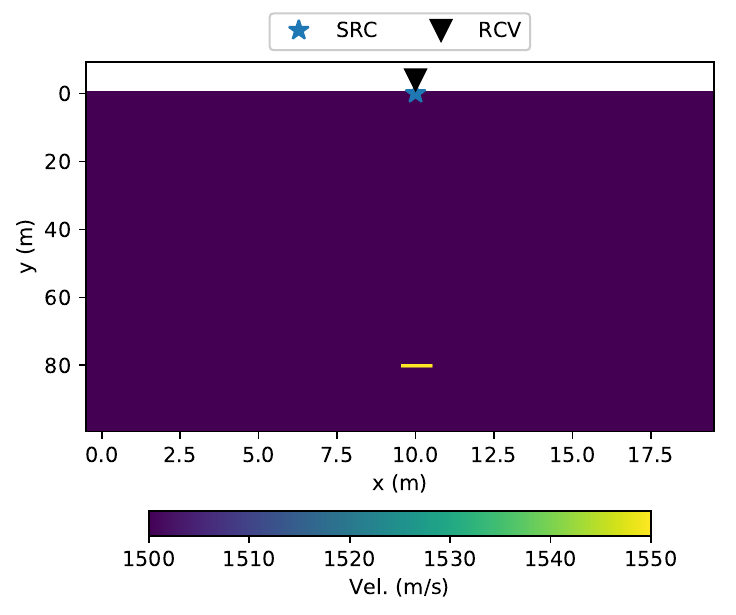}}
\subfloat[Scattering wavefield modelling. Direct wavefield was excluded in the modelling.\label{fig:rnn_2d_scattered_wavefield}]{\includegraphics[width=0.5\linewidth]{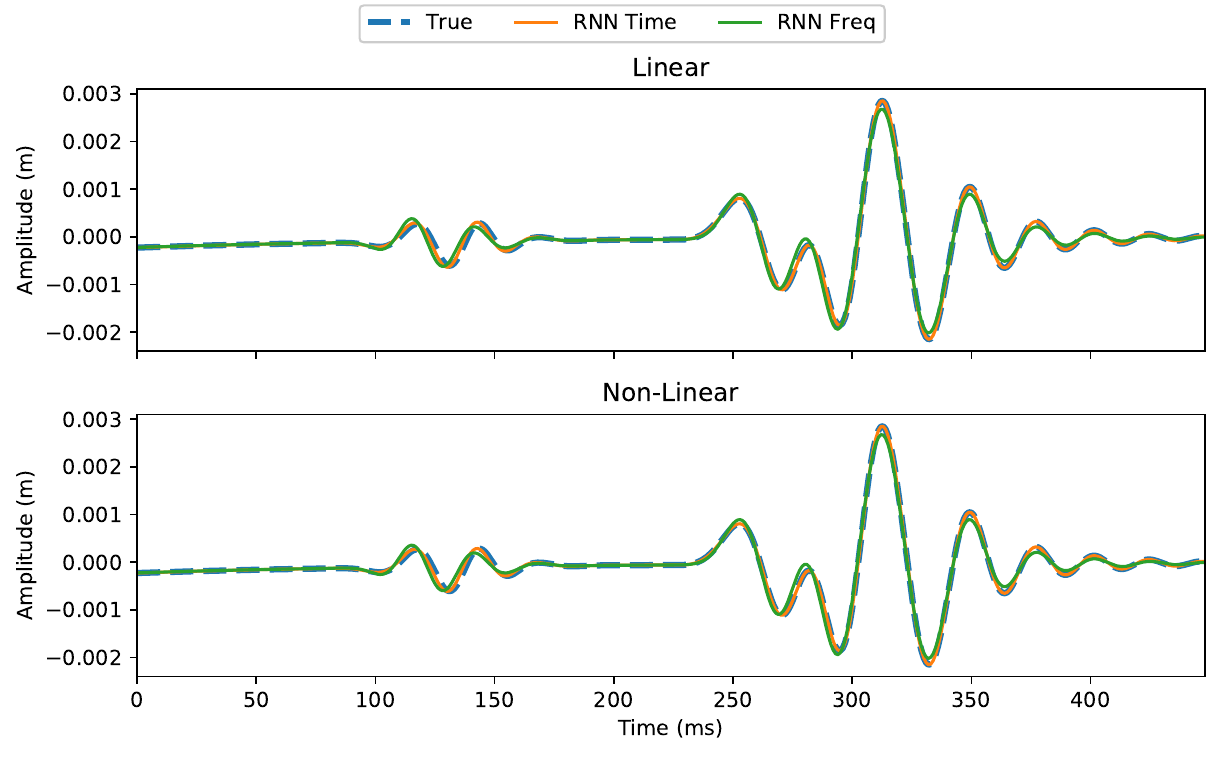}}
\caption[Scattering wave RNN forward modelling.]{Scattering wave RNN forward modelling.}    
\label{fig:rnn_2d_scattering}
\end{figure}

The remaining wavefield components are scattering waves. A constant velocity model of 1500 \si{ms^{-1}} was created with a 1550\si{ms^{-1}} point scatterer (Figure~\ref{fig:rnn_2d_scattered_vel}). RNN implementations were modelled to be depended non-linearly on the scattering amplitude and then approximately linearised. The results are given in Figure~\ref{fig:rnn_2d_scattered_wavefield}. The direct wave was not included in the scattered wavefield reconstruction. Similarly to previous components, scattering is modelled successfully.

Table~\ref{tab:rnn_2d_direct_multi} lists quantitative metrics for the wavefield components. RNN Freq was found to be better for imaging the direct wave (Table~\ref{tab:rnn_2d_direct_multi}), with an improvement of 0.01 in error tolerance and 0.3\% Relative Percentage Error (RPE). Metrics in Table~\ref{tab:rnn_2d_reflection} and Table~\ref{tab:rnn_2d_scattering} indicate that RNN Time matches the 2D Green’s function near perfectly, whilst RNN Freq introduce error of less than 0.04 and 0.1\% RPE. RNN Time is able to model the wavefield within a maximum 0.06 error tolerance and 1.74\% RPE, whilst RNN Freq is overall more accurate with 0.05 and 1.449\% respectively. Given these metrics and the observed models, the discrepancies between the analytical solution and the RNN implementation are deemed acceptable and should be suitable for the modelling process.

\begin{table*}[ht!]
    \footnotesize
    \centering
    % \ra{1.3}
    \subfloat[Direct wave.\label{tab:rnn_2d_direct_multi}]{
        \begin{tabular}{@{}lcc@{}}\toprule
            Modelling  & Error Tolerance    & RPE (\%) \\ \hline
            RNN Time   & 0.060 & 1.740            \\
            RNN Freq   & 0.050 & 1.449            \\ \hline
        \end{tabular}}\qquad
    \subfloat[Reflected and transmitted wave.\label{tab:rnn_2d_reflection}]{
        \begin{tabular}{@{}lcc@{}}\toprule
            Modelling  & Error Tolerance    & RPE (\%) \\ \hline
            RNN Time   & 0.001 & 0.002                  \\
            RNN Freq   & 0.020 & 0.013                  \\ \hline
        \end{tabular}}
    \\        
    \subfloat[Scattering wave.\label{tab:rnn_2d_scattering}]{
        \begin{tabular}{@{}lcc@{}}\toprule
            Modelling  & Error Tolerance    & RPE (\%) \\ \hline
            RNN Time – Non-linear	& 0.003	& 0.010     \\
            RNN Time – Linear   	& 0.010 & 0.025     \\
            RNN Freq – Non-linear	& 0.030	& 0.076     \\
            RNN Freq – Linear	    & 0.040 & 0.097     \\ \hline
    \end{tabular}}
    \caption[Empirical comparison of 2D wavefield components.]{Empirical comparison of 2D wavefield components.}
\end{table*}

\subsection{Gradient Comparison}
Classical FWI approaches generally use the adjoint state method to calculate gradients or the finite differences approach (although computationally expensive), whereas DNN frameworks use automatic differentiation. \cite{Richardson2018} already showed equivalence for the time-domain, and we now confirm the same for pseud-spectral RNN approach.

\begin{figure}[ht!]
    \centering
    \includegraphics[width=0.8\linewidth]{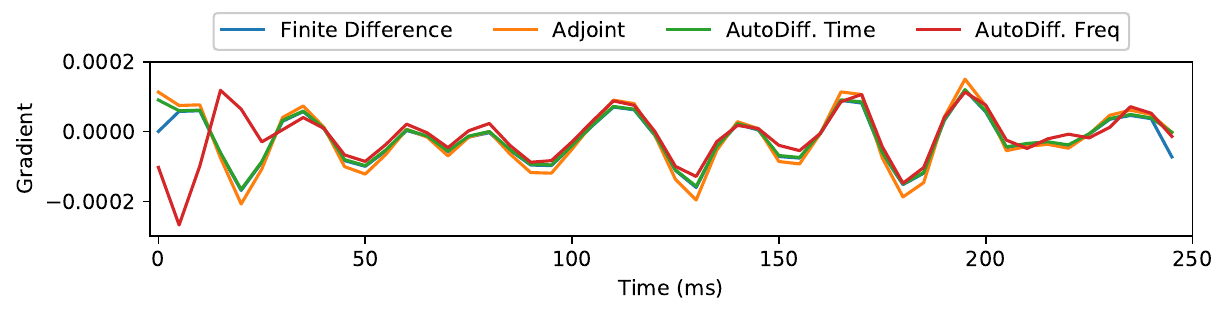}
    \caption[Gradient comparison of RNN implementation with classical approaches.]{Gradient comparison of of RNN implementation with classical approaches. AutoDiff. is the automatic differentiation implementation in Tensorflow v2.0.}
    \label{fig:rnn_gradient_comp}
\end{figure}

A random 1D model was generated, randomly perturbed and the gradient of the cost function was evaluated along the trace. Figure~\ref{fig:rnn_gradient_comp} and Table~\ref{tab:rnn_gradient_comp} compare the gradients at each point for classical finite differences and adjoint techniques to automatic differentiation (AutoDiff.). The adjoint state and AutoDiff. Freq reacts similarly and slightly overestimates the gradient, with the pseudo-spectral approach being worse. AutoDiff. Time underestimates the gradient with an infinitesimal error. Gradients deviate at the edges in either case, with AutoDiff. Freq producing evident perturbation in the initial few time-steps. This is due to the choice of the batch-size within the inversion process and is further discussed in subsection §~\ref{sec:rnn_hp_tunin}. Although this might seem worrying, the scale of this deviation is very minimal and no concerning effects were observed within the previous experimentation leading to this investigation. The other discrepancies are attributed to numerical inaccuracies as per \cite{Richardson2018}.
\begin{table*}[!ht]
    \footnotesize
    \centering
    % \ra{1.3}
    \begin{tabular}{@{}lccc@{}}\toprule
        \textbf{Finite Difference gradient baseline} & Adjoint & AutoDiff. Time & AutoDiff. Freq \\ \hline 
        Error tolerance                     & $1.000\times10^{-5}$ & $-2.196\times10^{-9}$      & $3.000\times10^{-4}$       \\
        RPE (\%)                            & $0.593$   & $1.302\times10^{-5}$      & $1.779$        \\ \hline 
    \end{tabular}
    \caption{Empirical comparison of gradient calculations.}\label{tab:rnn_gradient_comp}
\end{table*}

\subsection{Hyper-Parameter Tuning}\label{sec:rnn_hp_tunin}
\begin{figure}[ht!]
    \centering
    \includegraphics[width=0.95\linewidth]{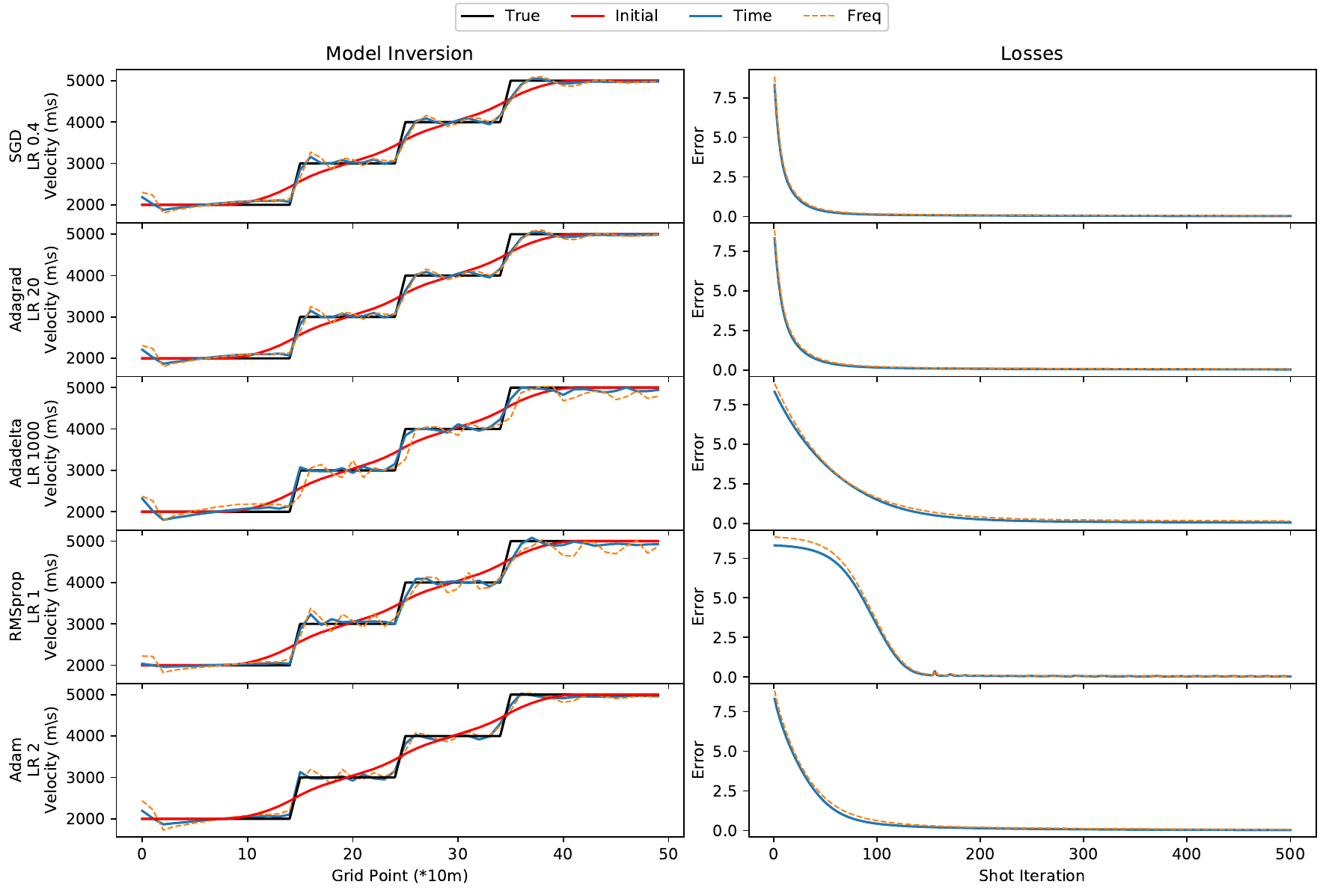}
    \caption[Hyper-Parameter tuning.]{Tuning of hyper-parameters to identify ideal loss optimizer combination.}
    \label{fig:rnn_hp_tuning}
\end{figure}

Similarly to the approach shown in \cite{Sun2019}, a benchmark 1D 4-layer synthetic profile, with velocities [2, 3, 4, 5]\si{kms^{-1}}, was used to identify the ideal parameters for the RNN architecture. This is illustrated as the Black line in Figure~\ref{fig:rnn_hp_tuning}. Classical 1D second-order finit-difference modelling was used to generate the required true receiver data. Multiple learning rates for the different loss optimizers were investigated to try and identify the ideal combination. Figure~\ref{fig:rnn_hp_tuning} shows the best combination for all losses with an ideal batch size of three. The full investigation for this tuning is given in Appendix~\ref{sec:app_results_rnn_hp_tuning}. 

Left side of Figure~\ref{fig:rnn_hp_tuning} shows the inverted velocity profiles, with Red being the initial velocity profile. For Stochastic Gradient Descent, the learning rates was found to be both between zero and one. This is as expected and follows conventional loss optimization. On the other hand, the other loss optimizers had to be scaled to beyond one due to the magnitude differences brought by accumulated squared-norms of the gradients as investigated by \cite{Sun2018}. This is allowed provided the scaling coefficient is between zero and one. For Adagrad, following from \cite{Duchi2011}, the $\beta$ hyper-parameter was fixed at 0.9 and learning rate found to be 20. Adadelta, RMSprop and Adam optimal learning rates were identified at 1000, 1 and 2 respectively. 

The right side of Figure~\ref{fig:rnn_hp_tuning} gives the loss progression. All optimizers iteratively reduce the error with additional shots and on similar scales. Stochastic Gradient Descent and Adagrad do this relatively sooner than the rest, yet the inverted velocity is not as good as the other optimizers. RMSprop follows a rather slow gradual decrease in loss, which then sudden increases. This is expected given that RMSprop updates are derived from a moving average of the square gradients and require an inertial start.

\begin{figure}[ht!]
    \centering
    \includegraphics[width=0.8\linewidth]{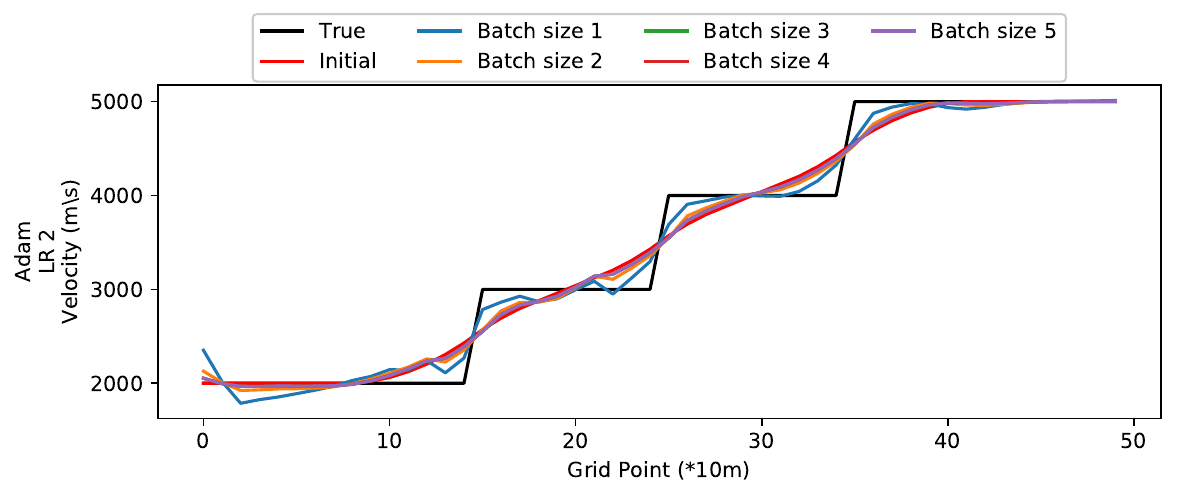}
    \caption[Effect of batch-size on inversion process.]{A smaller batch-size introduces error at the initial part of the velocity profile due to more localized updates. This was derived for a time time implementation for RNN architecture with Adam loss optimizer and learning rate of 2.}
    \label{fig:rnn_hp_tuning_batch_size}
\end{figure}

Based on this investigation, \textbf{Adam} with a learning rate of 2 was identified as the best optimizer. This provided the most stable inversion for either RNN Time or Freq, with the most update and reasonable error loss performance. Mis-match in the shallow part of the velocity is due to the choice of batch-size within the RNN update process. Figure~\ref{fig:rnn_hp_tuning_batch_size} shows the Adam optimizer fixed with learning rate 2 and inverted for batch sizes ranging from one to five. The smaller the batch size, the greater the error since the inversion is more localized and amplifies the gradient onset error shown in Figure~\ref{fig:rnn_gradient_comp}. The larger the batch size, the better the inversion as more data is being used. This poses a limitation since batch size is limited by the Graphical Processing Unit RAM. Given fore-sight that this approach will be used on a large dataset, this was taken as a caveat and batch size fixed at one for the rest of the implementation. 

\subsection{Marmousi Model Experiment}\label{sec:results_marmousi_dataset}
\begin{figure}[ht!]
    \centering
    \includegraphics[width=0.98\linewidth]{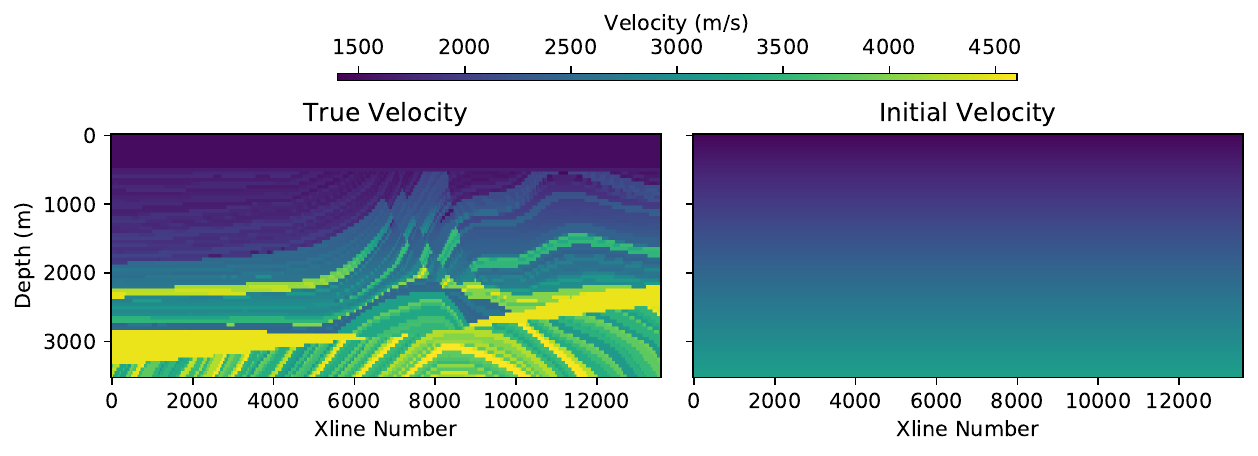}
    \caption[Synthetic 2D Marmousi models for RNN training.]{50\si{m}$\times$50\si{m} grid 2D Marmousi models for RNN training.}
    \label{fig:rnn_marm_models}
\end{figure}

The Marmousi-2 model \cite{Martin2002} was used to evaluate theory-guided RNN on an industry-standard dataset. This was re-sampled to a 50\si{m}$\times$50\si{m} grid and smoothed to create the initial model. These velocity models are plotted in Figure~\ref{fig:rnn_marm_models}. True synthetic receivers were computed by forward modelling through the RNN framework. 56 shots at 300m intervals at depth 200m were generated with a Perfectly Matched Layer at the boundaries. Receivers were set at 50m intervals and modelled for 12\si{s} duration.

\subsubsection{Training of RNN}
\begin{figure}[ht!]
    \centering
    \includegraphics[width=0.8\linewidth]{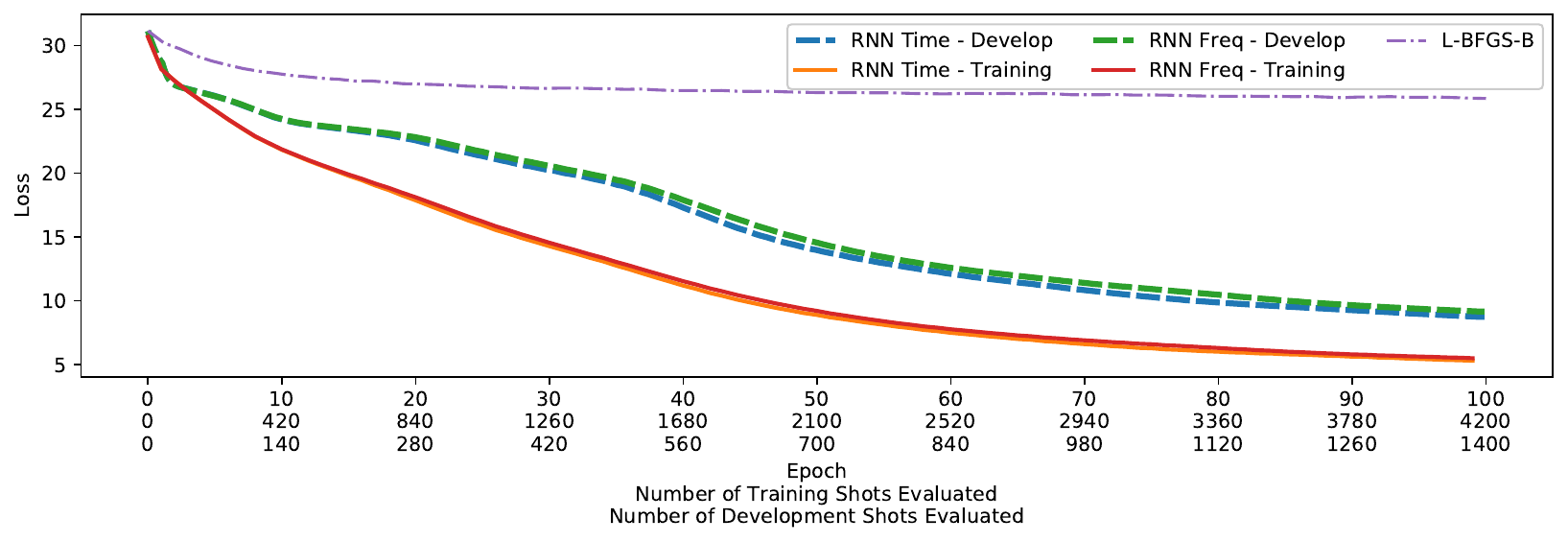}
    \caption[RNN loss performance.]{RNN loss performance for RNN training and development datasets using Adam optimizer with a learning rate of 2 and batch size 1. The horizontal labels show the epoch number and respective number of shots evaluated for training and development. L-BFGS-B is the cost function evaluation for classical FWI plotted on shot number equivalent. Either RNN approach converges quicker than L-BFGS-B, and RNN Freq provides a more stable convergence and better performance than RNN Time.}
    \label{fig:rnn_losses}
\end{figure}

As in standard RNN approaches, the receiver dataset was split into training and development datasets with a 75\%-25\% split. The training was run for 100 epochs, with early stopping on an NVIDIA Titan V Graphical Processing Unit courtesy of Istituto Nazionale di Geofisica e Vulcanologia. Development loss was calculated every 5th training shot. Figure~\ref{fig:rnn_losses} gives the RNN performance for training and development datasets using Adam optimizer with learning rate of 2.0 and batch size 1. The horizontal labels shows the epoch number and respective number of shots evaluated for training and development. Computational run times are of 14 hours per approach. Both RNN Time and RNN Freq follow similar reductions in loss per epoch and indicate that either implementation converge to an optimal loss. L-BFGS-B loss for classical FWI is shown and is discussed is in the next section.

\subsubsection{Comparison with classical FWI}\label{sec:results_rnn_comparison_to_FWI}

3.5\si{Hz} FWI with Sobolev space norm regularization \cite{Kazei2019} was used to compare against theory-guided FWI. This results in a minimum update resolution of 414m, and the iterative update process started from frequency 1\si{Hz} and iteratively updated by a factor of 1.2. The optimization algorithm was L-BFGS-B, with 50 iterations per frequency band in each update. Forward shot modelling was done every 100m, starting from 100m offset, and receivers spaced every 100m. 

Figure~\ref{fig:rnn_losses} plots the cost function versus the number of shot evaluation equivalent for classical FWI and RNN. The RNN framework is more computationally efficient since either RNN approach converge significantly quicker than L-BFGS-B. RNN Freq provides a more stable convergence and is better performant then RNN Time. The classical FWI is plotted as a shot number equivalent and not the epoch number. The full cost function performance is provided in Appendix~\ref{sec:app_results_classical_FWI}. 

Figure~\ref{fig:rnn_models} compares the inverted velocities and residuals for FWI, together with RNN Time and RNN Freq implementations. Complementary plots showing the model update progressions for this section are provided as part of Appendix~\ref{sec:app_results_rnn_update_progress}. The true model velocity in Figure~\ref{fig:rnn_models} identifies three zoomed areas which are shown in Figure~\ref{fig:rnn_model_zoomed_in} and Figure~\ref{fig:rnn_velocity_profiles} are velocity profiles taken at 2000 Xline intervals. Figure~\ref{fig:rnn_model_spectra} show the resolution spectra derived via FFT on the velocity models. Comparing FWI and the RNN model in either of these figures, it is clear that the resolution recovery is different. Figure~\ref{fig:rnn_model_spectra} confirms the frequency content in these approaches and shows how RNN models invert more of the lower frequencies in Zoom 2 and Zoom 3. In Zoom 1, FWI is slightly better at frequency recovery beyond 25Hz.

Residual plots (Figure~\ref{fig:rnn_models}-{~\ref{fig:rnn_model_zoomed_in}) and the velocity profiles (Figure~\ref{fig:rnn_velocity_profiles}) show how RNN approaches are able to recover more of the signal in the shallow right side (Zoom 1) and the over-thrust middle area (Zoom 2) of the model. Almost all the signal up to depth 1500m is inverted correctly in Zoom 1 whereas over-thrust faults are near perfectly recovered in Zoom 2 and ~\ref{fig:rnn_model_zoomed_in}F. Zoom 3 is of most interest. The prominent layer at depth circa 2000m is nearly completely missed by RNN models, whereas FWI is able to recover this partially. On the other hand, the deeper 3000m strata are hardly identified with FWI. Residual figures in the full sections show that the RNN model amplitude recover is not as good when compared to FWI (Labels~\ref{fig:rnn_model_zoomed_in}A-B). Indeed, some layers are missed at depth greater than 1500m for Xline number greater than 10,000 (Label~\ref{fig:rnn_model_zoomed_in}C-D). Considering either RNN approach in Figure~\ref{fig:rnn_models}, there is a low-frequency \textit{shadow} artefact introduced till depth 2300m from Xline 0 to 6000 and Xline 8000 to 13900. This is attributed to the practical implementation of batch-size discussed in §~\ref{sec:app_results_rnn_hp_tuning}.

Figure~\ref{fig:rnn_receivers} shows labelled receivers for either model at CDP 60, 150 and 300. These CDPs split the model into three sections, representing the different extremities. Label A and B reiterate that the shallow left side is better imaged for FWI, whilst shallow right side is better for RNNs respectively. Label C is the missing high velocity at depth 2000m which has incorrect amplitude for the RNNs, but positioned correctly. Classical FWI has less prominent leakage in this area, yet very evident. Label D is the badly imaged layer at depth between 2000m and 2500m on the right side of the model. Labels E throughout the residuals highlight better low frequency resolution imaging by RNN approaches. Indeed, RNN Freq is able to recover slightly more of these low frequencies and identified by E$^{1}$ and E$^{2}$. Similar improvements are visible throughout the other plots. 

\begin{figure}[ht!]
    \centering
    \includegraphics[width=0.99\linewidth]{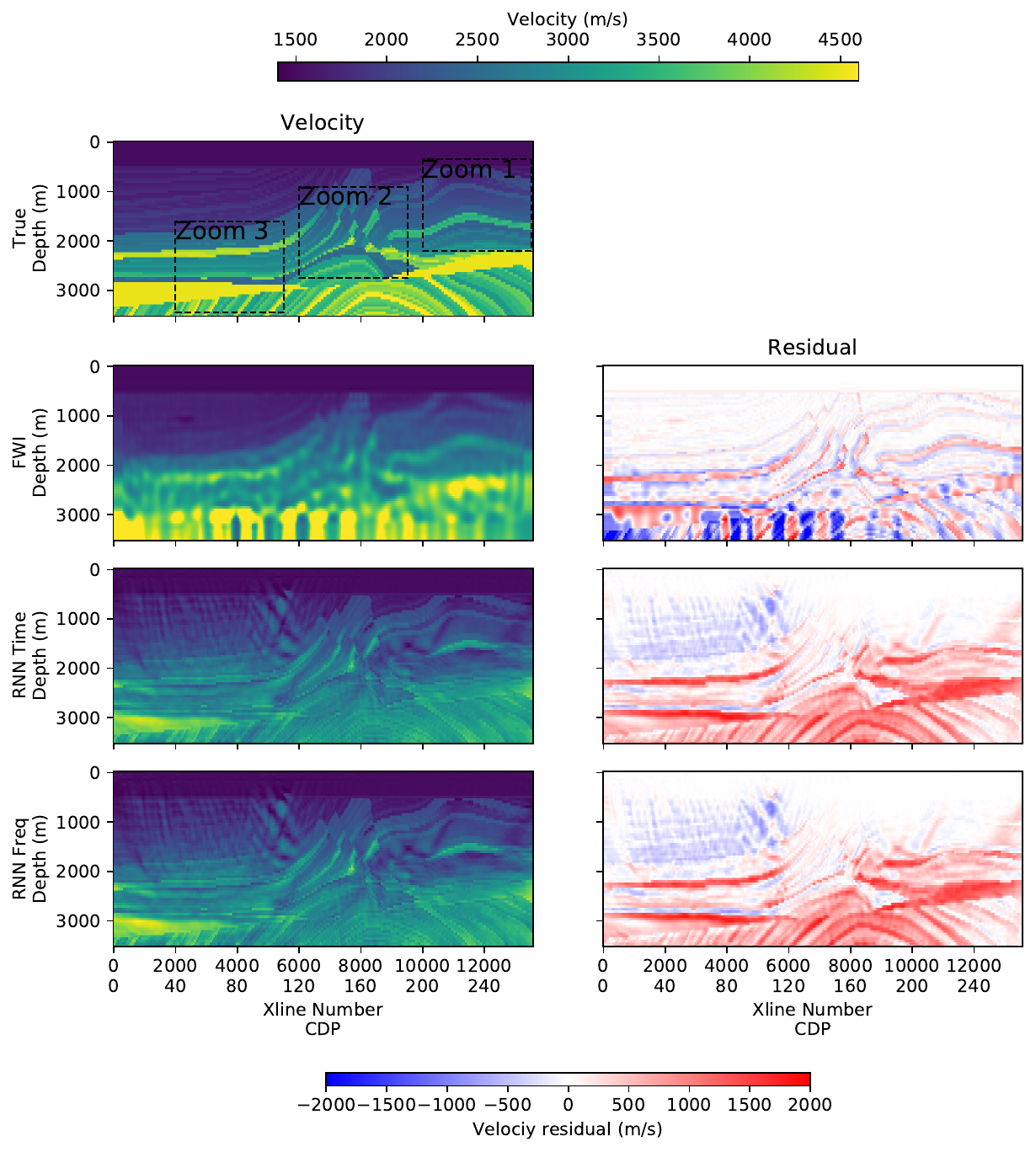}
    \caption[Classical FWI and RNN implementation velocity model inversion.]{Classical FWI and RNN implementation velocity model inversion.}
    \label{fig:rnn_models}
\end{figure}

\begin{figure}[ht!]
    \centering
    \includegraphics[width=0.99\linewidth]{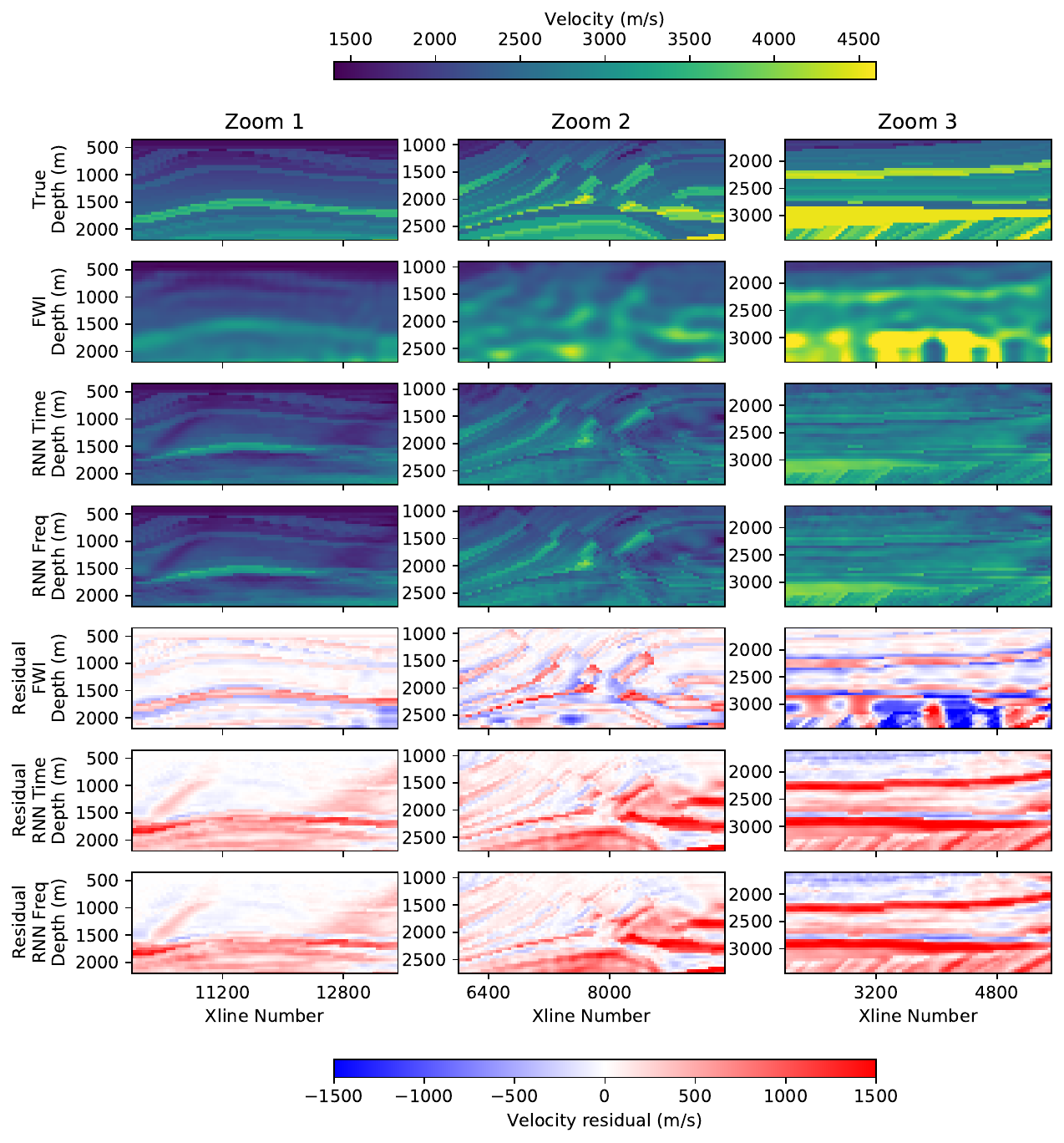}
    \caption[Zoomed In RNN Models]{Zoomed In RNN Models}
    \label{fig:rnn_model_zoomed_in}
\end{figure}

\begin{figure}[ht!]
    \centering
    \includegraphics[width=0.8\linewidth]{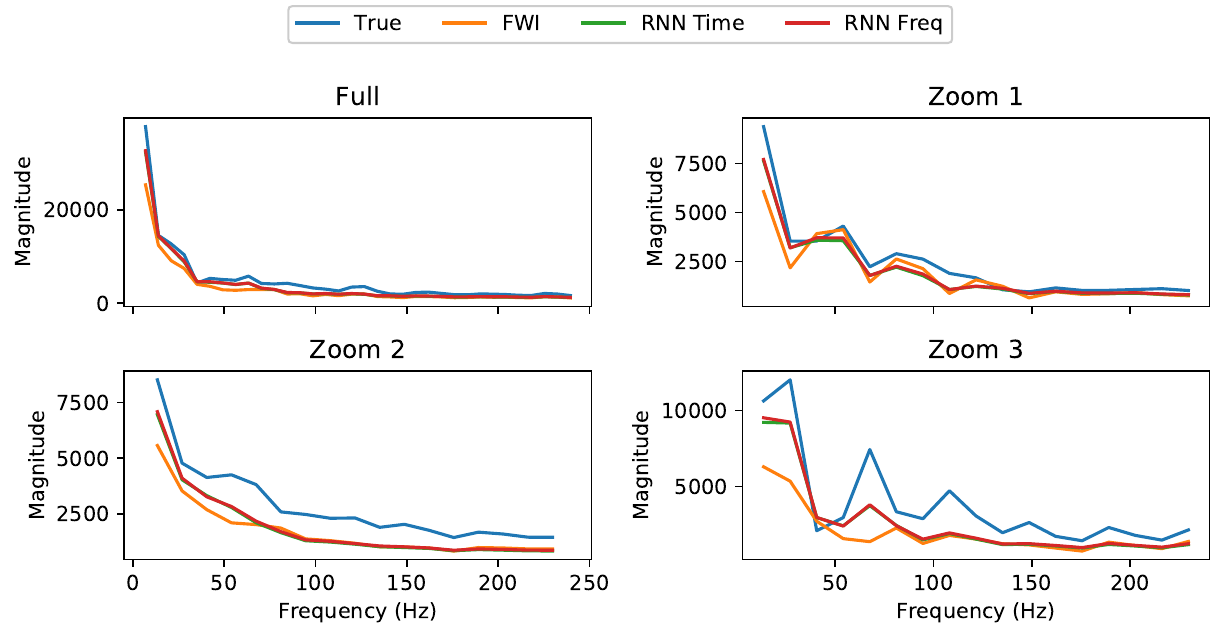}
    \caption[RNN models velocity resolution spectra]{RNN model velocity resolution spectra.}
    \label{fig:rnn_model_spectra}
\end{figure}

\begin{figure}[ht!]
    \centering
    \includegraphics[width=0.8\linewidth]{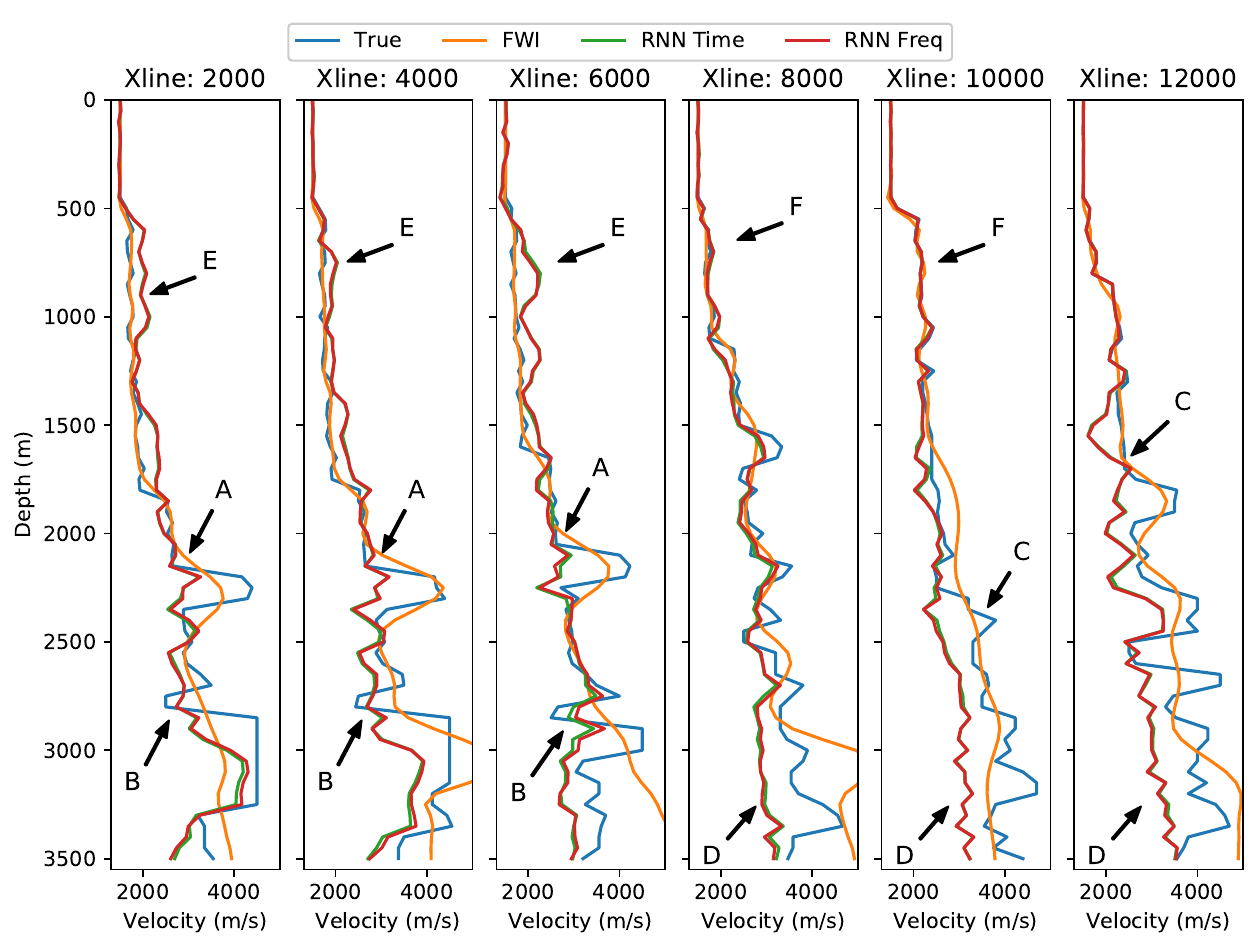}
    \caption[Velocity profiles for RNN and classical FWI.]{Comparison of velocity profiles for RNN and classical FWI. Label \textbf{A-B}: RNN is able to identify strata near perfectly, however unable to inverte the amplitudes values correctly. Label \textbf{C-D}: Missed layers from RNN approaches. Label \textbf{E}: Low frequency artefact for RNN. Label \textbf{F}: Near perfect velocity inversion in the middle Xlines, over shallow depth.}
    \label{fig:rnn_velocity_profiles}
\end{figure}

\begin{figure}[ht!]
    \centering
    \includegraphics[width=0.8\linewidth]{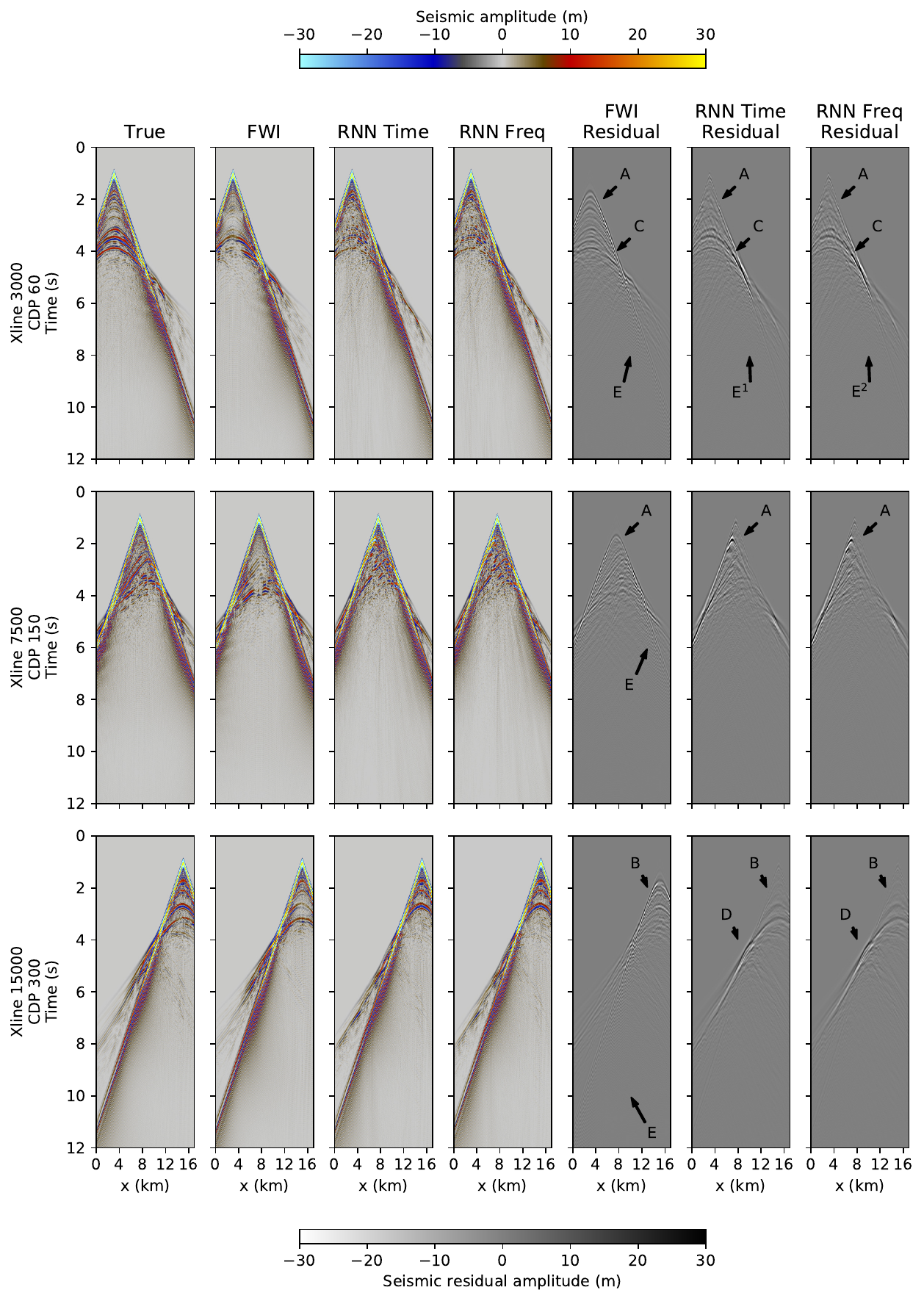}
    \caption[Labelled receivers for True, FWI and RNN models.]{Receivers for True, FWI and RNN models at CDP 60, 150 and 300. Label \textbf{A} and \textbf{B}: Shallow left side is better imaged for FWI. Label \textbf{C}: Missing high velocity with incorrect amplitude but positioned correctly. Label \textbf{D}: Badly imaged layer. Labels \textbf{E}: Better low-frequency imaging for either RNN approaches. Labels \textbf{E$^{1}$} and \textbf{E$^{2}$}: RNN Freq is able to recover slightly more low frequencies. 
    }
    \label{fig:rnn_receivers}
\end{figure}

\section{Discussion}\label{sec:discussion}
\subsection{Inversion Paradigm}
Theory-guided inversion inherits advantages and problems from either inverse theory and deep learning. It faces challenges of cycle-skipping and local-minima, whilst it benefits from the use of automatic differentiation to calculate the gradient. This reduces development time as it avoids the need to manually implement and verify the adjoint state method. Furthermore, being at the intersection of physics and computer science, it is inherently strengthened by contributions from two communities of researchers. This opens up the possibility of considering other deep learning techniques such as dropout or other acyclic-graph architectures such as directed acyclic graphs \cite{Bogaerts2020}.

In classical FWI, the wavefield at the final time step is affected by the wavefield during the initial time steps. Back-propagation must occur over the entire sequence of time steps for theory-guided RNN. Application of back-propagation through thousands of layers is not a typical application in deep learning applications and automatic differentiation is not designed to efficiently handle such situations. Strategies common to other FWI frameworks to reduce memory requirements could be translated into the field. Examples would include not saving the wavefield at every time step \cite{Nguyen2015}, applying compression to the wavefield time slices \cite{Boehm2015,Kalita2017}, saving wavefields to disk rather then memory \cite{Shen2015}, and regenerating wavefields during back-propagation rather than storing them \cite{Malcolm2016, Yao2020}. 

\subsection{Training Datasets for Real Data}\label{ref:sec_disc_development_dataset}
For theory-guided RNN, excluding part of the data from training for use as a development dataset is standard practice in deep learning, but not within classical FWI. For a real-world problem, the size of the seismic dataset relative to the model parameters generally has fewer data samples and could potentially prove problematic. Hyper-parameter tuning for the optimal parameters for RNN demonstrate that practice can result in convergence to a good model, yet this does not prove a similar result is achievable when using the entire dataset. 

\subsection{Forward Modelling and Multiples}
All shots considered within the forward problem for either FWI and RNN framework were within the water column for the Marmousi model. This implies that receiver data have surface-related multiples, together with all other inter-layer multiple components. Undergoing forward problem solving with and without multiples is a decision that the literature is still unable to resolve. 

Multiples travel longer paths and are reflected at small angles in contrast to the primaries and are able to illuminate shadow zones where primary reflections cannot reach \cite{Bergen2019}. The inclusion of these wavefield components can lead to improvement within the inversion process as multiples can contain more subsurface information compared to primary and diving wave \cite{Komatitsch2002}. \cite{Bleibinhaus2009} investigated these effects of surface scattering in FWI and concluded that velocity models resulting from neglecting the free surface in the inversion show artefacts and suffered from a loss of resolution. \cite{Liu2020} employ a combination of lower-order multiple as the source and the higher-order multiple to invert, whilst \cite{Zhang2013} transform each hydrophone into a virtual point source with a time history equal to that of the recorded data to help their inversion and are able to produce methods utilizing multiples to improve velocity updates. 

\cite{Hicks2001} and \cite{Operto2006} show how traditional FWI would become unstable when inverting with free surface-related waves. This said, removing multiples introduce additional processing steps which are subject to error and could lead to the removal of the signal. The consensus is that the choice of multiple inclusion is per different use cases. Indeed, the work presented could be revisited for the sensitivity of multiples within the inversion.

\subsection{Implications of Data Volume and Computational Power}
More data is directly correlated with better modelling for NN frameworks, and this ability is limited by the resources available. Similar to classical FWI, computational power is a limitation within the frameworks presented. This was already identified within the RNN approach with the limit from the Graphical Processing Unit RAM, constraining the model size and batch processing. A larger batch-size for RNN processing would intuitively imply that the optimization is less likely to get stuck with local minimum and reduce the probability of cycle-skipping. Workaround for this could be multi-Graphical Processing Unit systems, such as NVIDIA's DGX station\footnote{\url{https://www.nvidia.com/en-us/data-center/dgx-systems/}$\textsuperscript{\ref{ref:footnote_no_affiliation}}$} and  Lambda Lab's Vector station\footnote{\url{https://lambdalabs.com/gpu-workstations/}$\textsuperscript{\ref{ref:footnote_no_affiliation}}$}, or cloud computing such as Amazon Web Services \footnote{\url{https://aws.amazon.com/nvidia/}$\textsuperscript{\ref{ref:footnote_no_affiliation}}$} and Google Cloud\footnote{\url{https://cloud.google.com/gpu/}$\textsuperscript{\ref{ref:footnote_no_affiliation}}$}\footnote{\label{ref:footnote_no_affiliation}These are just samples of resources and there is no affiliation.}. This cost on memory requirements for NN is a common issue with solving optimization of large-scale neural networks \cite{Bottou2018} and efforts have been made into mitigating this via alternating gradient direction methods and Bregman iteration training methods \cite{Boyd2011, Taylor2016}.

\section{Conclusion}\label{sec:conclusion}
In this manuscript, a theory-guided approach for FWI using RNN was derived. This was developed theoretically, qualitatively assessed on synthetic data and tested on the Marmousi dataset. Theory-guided RNN as an analogue of FWI was implemented for 2D experiments and different wavefield components compared to an analytical 2D Green's function and time implementation. Based on these results, RNN Time is able to model the wavefield within a maximum 0.06 error tolerance and 1.74\% RPE. RNN Freq is overall more accurate with 0.05 error tolerance and 1.449\% RPE. Assessment on the gradients indicates how the adjoint state and RNN Freq gradients in general overestimate finite difference calculation, whilst RNN Time under-estimates it with an infinitesimal error. RNN Freq produced a perturbation on the onset of the gradient which was attributed to modelling artefact and could be mitigated in future versions of this approach. Based on the model size and compute available, the ideal loss was Adam with a learning rate of 2 and batch size of 1. Model batch size proved to be a limitation for practical implementations, yet RNN is computationally more efficient than the classical FWI presented in this work. RNN freq provides more stable convergence and is better performant. Overall, RNN frameworks are able to identify faults, but amplitudes are not fully inverted properly.

RNN approach benefits from the wider community of active researchers. The reduction in development time is a direct integration from Computer Science to geophysics. Vice-versa, Deep Learning frameworks can adopt strategies common to FWI. 

The forward modelling approach used through this work was critiqued for the use of multiples. Whether to use or not to use multiples within forward modelling is model dependent and should be evaluated for RNN Freq. Similar to classical FWI, computational power was identifiable as a limitation within these DNN frameworks. Although this is currently a limitation, it will not be in the near future due to the relative quick development of GPUs. A corollary to the whole approach was addressed in the form of the maturity of the approach. 35 years of advances applied to these frameworks would be expected to yield very good results. Finally, other areas of DNN that can be applied to FWI were presented. Alternative architectures such as Transformers and use of Fourier Recurrent Units are readily available. Potential of transfer learning and solving differential equations using NN were presented as future directions of research for these frameworks.

\appendix
\section{LSTM Components}\label{sec:app_theory_Individual_RNN_Components}
\subsection{Forget gate}
The forget gate uses a sigmoid function to decide what information should be passed between hidden states. Values from this gate range between 0 and 1, indicating the level of information to be forgotten.

\subsection{Input gate}
The input gate uses a sigmoid activation function, accepts the previous hidden state and current input and decides which values will be updated. The current input and previous hidden state are passed into the tanh function to squeeze values between -1 and 1 and get a potential new candidate.
 	
\subsection{Cell state}
The cell state acts as a mechanism to transfers information through the sequence. This enables information from earlier time steps to be available at later time steps, thus reducing the effects of vanishing gradient. The preservation of gradient information by LSTM is illustrated in Figure~\ref{fig:LSTM_gradient}.

\begin{figure}[ht!]
    \centering
    \includegraphics[width=0.8\linewidth]{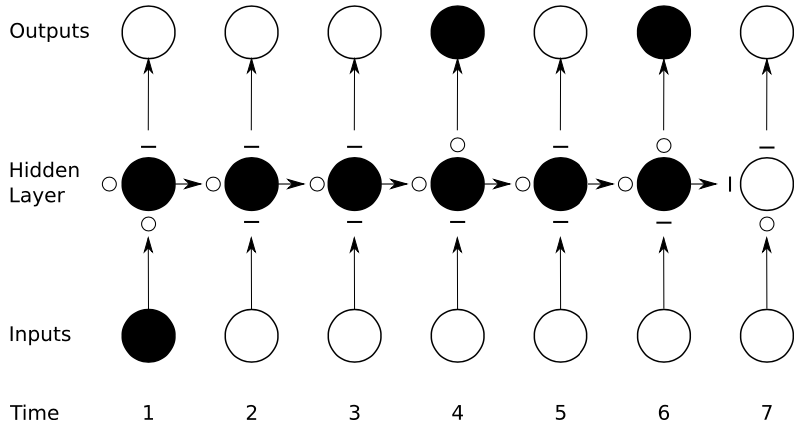}
    \caption[LSTM preserves the gradient to deeper layers of a NN.]{``Preservation of gradient with LSTM structure. The shading of the nodes indicate the influence of the inputs at a particular point in time. The black nodes indicate maximum sensitive and the white nodes are entirely insensitive. The state of the input, forget, and output gates are displayed below, to the left and above the hidden layer respectively. In this example, all gates are either entirely open (‘O’) or closed (‘—’)." From \cite{Graves2012}.}
    \label{fig:LSTM_gradient}
\end{figure}     
 
\subsection{Output gate}
The output gate determines the next hidden state. It 
 uses a sigmoid activation on the current state and previous hidden state, and multiples this new cell state with a tanh to decide which part of the data should be pushed forward through the sequence.

\subsection{RNN Hyper-Parameter Tuning}\label{sec:app_results_rnn_hp_tuning}
Similarly to the approach shown in \cite{Sun2019}, a benchmark 1D 4-layer synthetic profile, with velocities [2, 3, 4, 5] \si{kms^{-1}}, was used to identify the ideal parameters for the RNN architecture. Classical 1D second-order FD modelling was used to generate the required true receiver data. Batch size is used as a discriminator throughout Figure~\ref{fig:app_results_rnn_hp_tuning_inversion}. The results indicate that the larger the batch size used, the better the inversion as more data is being used. However, given fore-sight that this hyper-parameter tuning will be used a large dataset that might not fit in Graphical Processing Unit RAM, this was fixed at batch size one. 

\begin{figure}[ht!]
        \centering
        \includegraphics[width=0.99\linewidth]{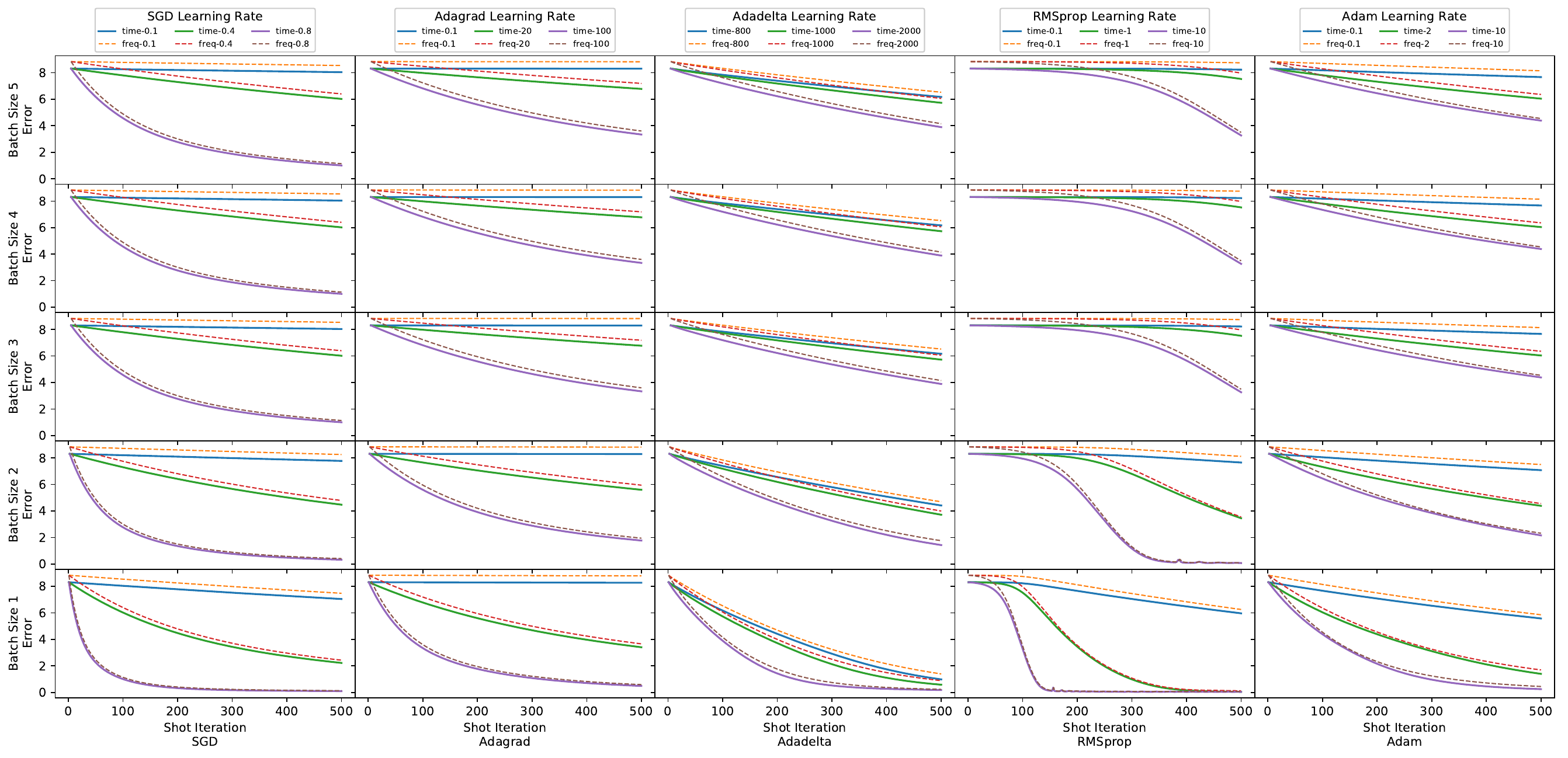}
        \caption[Losses for different loss optimizer learning rate hyper-parameter tuning.]{Losses for different loss optimizer learning rate hyper-parameter tuning.}
	\label{fig:app_results_rnn_hp_tuning_losses}
\end{figure}

\begin{figure}[ht!]
	\centering
	\subfloat[RNN Time]{\includegraphics[width=0.95\linewidth]{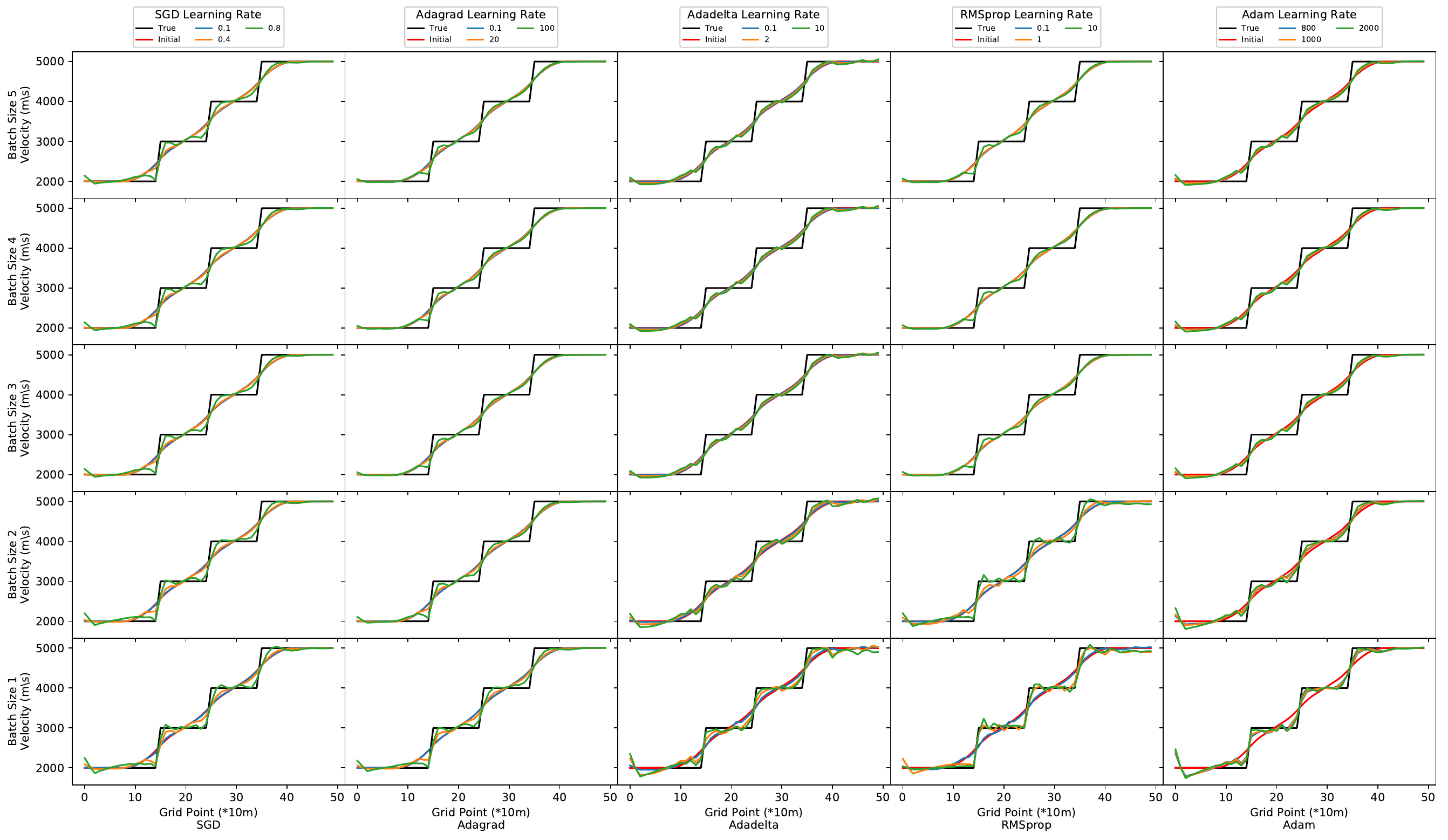}}
        \\
        \subfloat[RNN Freq]{\includegraphics[width=0.95\linewidth]{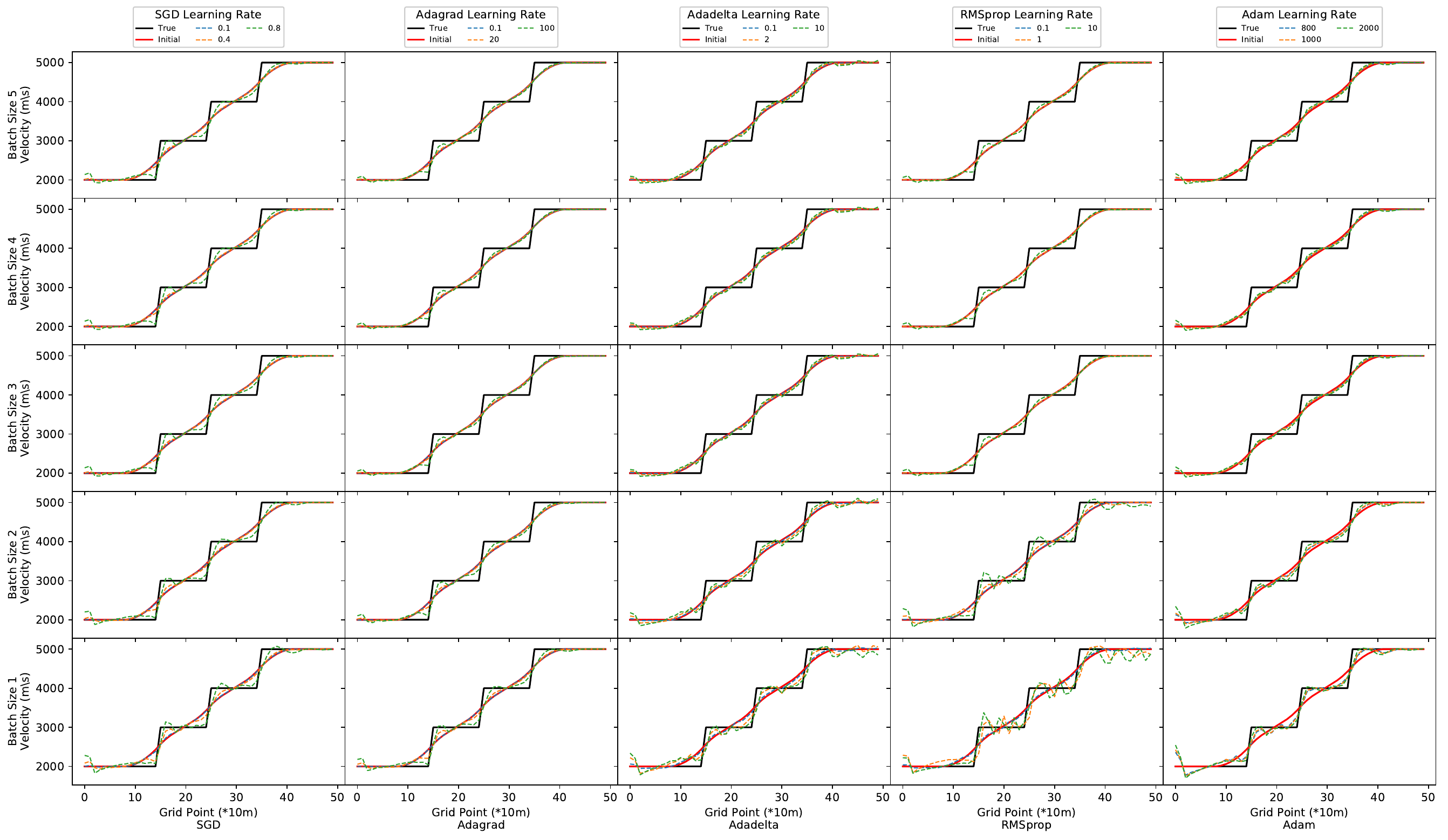}}
	\caption[Different Loss optimizer learning rate hyper-parameter tuning results.]{Loss optimizer learning rate hyper-parameter tuning results.}
	\label{fig:app_results_rnn_hp_tuning_inversion}
\end{figure}

\subsection{RNN Inversion Update Progress}\label{sec:app_results_rnn_update_progress}
Complementary to inverted Marmousi models in §~\ref{sec:results_rnn_comparison_to_FWI}, Figure~\ref{fig:app_results_rnn_update_progress_model} gives the update progress at epoch 10, 25, 40, 55, 70, 85 and 100 for RNN Time and RNN Freq, together with residual. Furthermore, classical FWI progress is included at different update frequency scales. In addition, receivers are provided in Figure~\ref{fig:app_results_rnn_update_progress_rcv}.

\begin{figure}[ht!]
        \centering
        \includegraphics[width=0.8\linewidth]{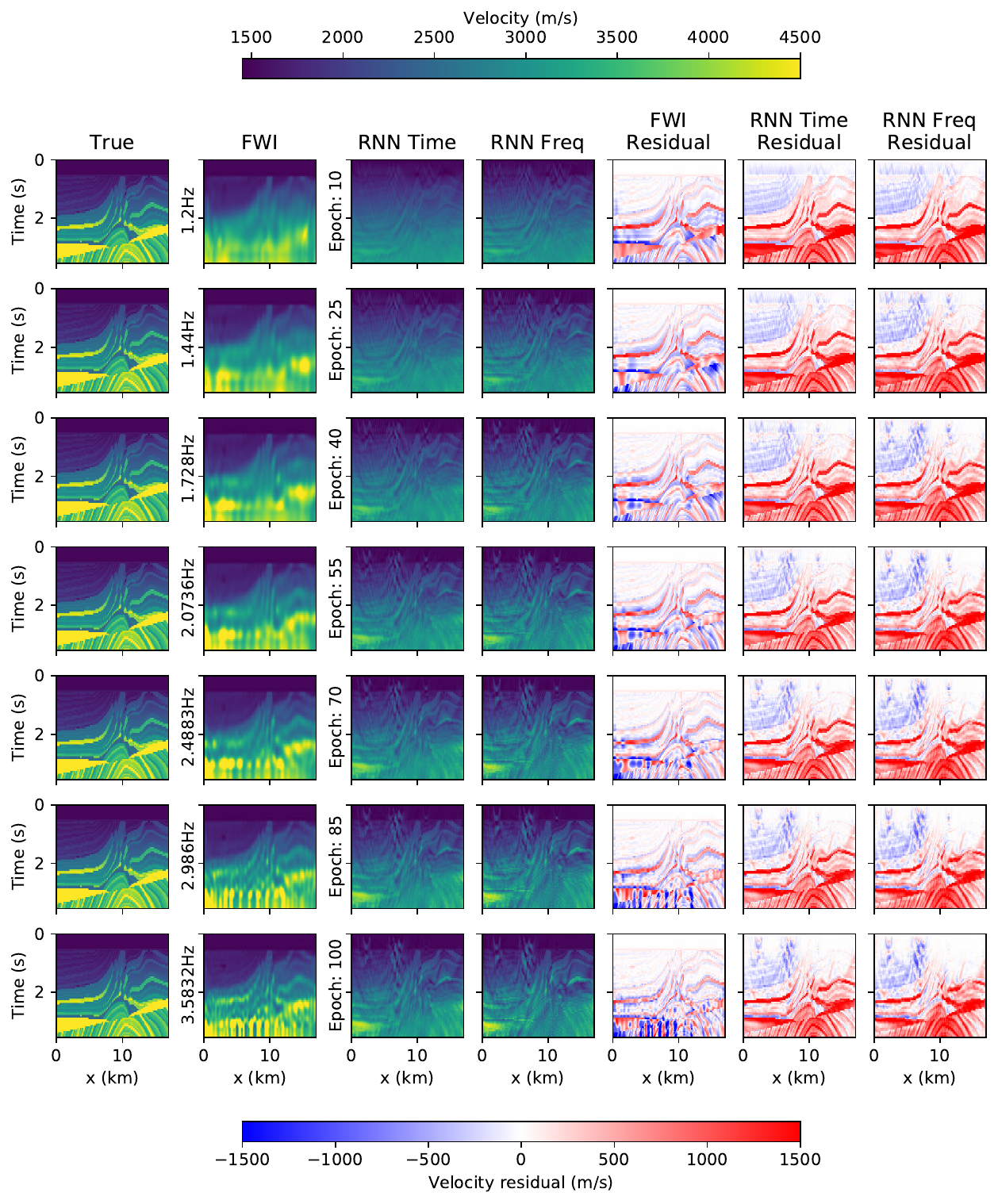}
        \caption[Velocity model inversion update progress.]{Velocity model inversion update progress for classical FWI, RNN Time and Freq, with residuals.}
	\label{fig:app_results_rnn_update_progress_model}
\end{figure}

\begin{figure}[ht!]
        \centering
        \includegraphics[width=0.8\linewidth]{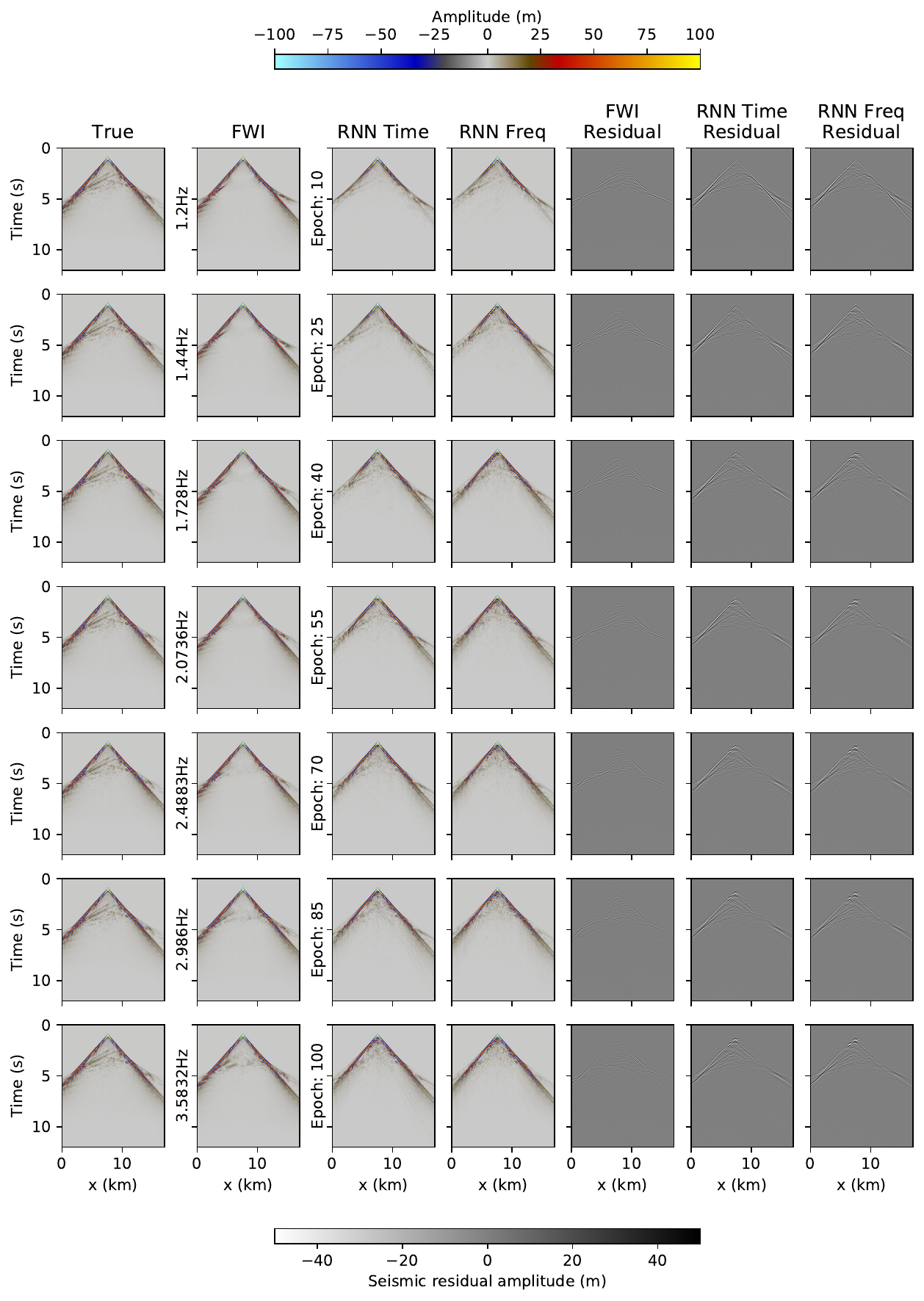}
        \caption[Receiver progress through model updates.]{Receiver progress through model updates for classical FWI, RNN Time and Freq, with residuals.}
	\label{fig:app_results_rnn_update_progress_rcv}
\end{figure}

\section{Classical FWI}\label{sec:app_results_classical_FWI}
\subsection{Inversion}
FWI with Sobolev space norm regularization was used as the deterministic version of FWI within this work. The maximum frequency of the inversion process was set to be 3.5\si{Hz}. The iterative update process started from frequency 1\si{Hz} and iteratively updated by a factor of 1.2 until reaching a maximum frequency of 3.45\si{Hz}. The optimization algorithm was L-BFGS-B, with 50 iterations per frequency. Figure~\ref{fig:app_classical_fwi_loss} is the loss update for L-BFGS-B and Stochastic Gradient Descent. Figure~\ref{fig:classical_fwi_progression} shows the progression of the frequency updates.

\begin{figure}[ht!]
        \centering
        \includegraphics[width=0.8\linewidth]{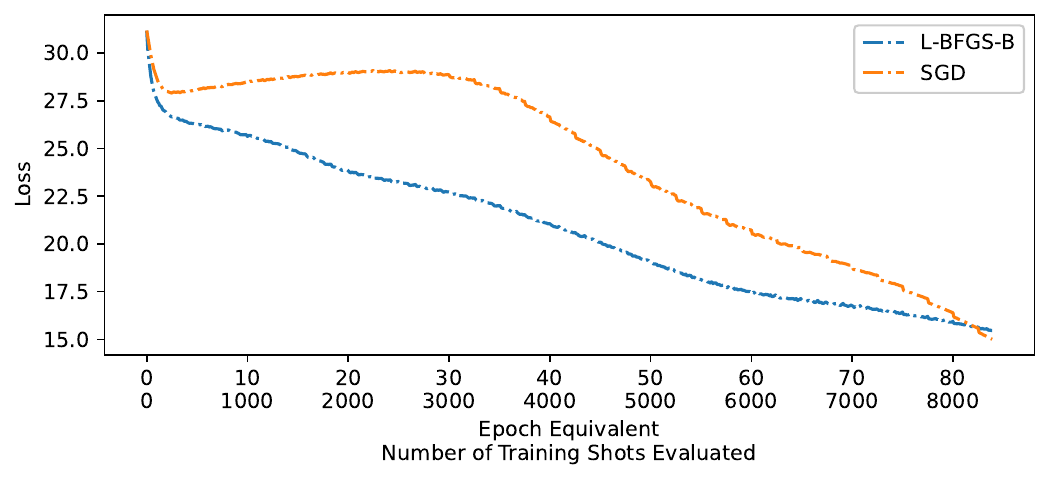}
        \caption[Classical FWI loss update.]{Classical FWI loss update for L-BFGS-B and Stochastic Gradient Descent. L-BFGS-B was a better loss optimizer than Stochastic Gradient Descent due to the monotonically decreasing loss. Stochastic Gradient Descent training should have been stopped at an earlier epoch due to the increase at 30 when compared to earlier epoches.}
        \label{fig:app_classical_fwi_loss}
\end{figure}

\begin{figure}[ht!]
    \centering
    \includegraphics[width=0.44\linewidth]{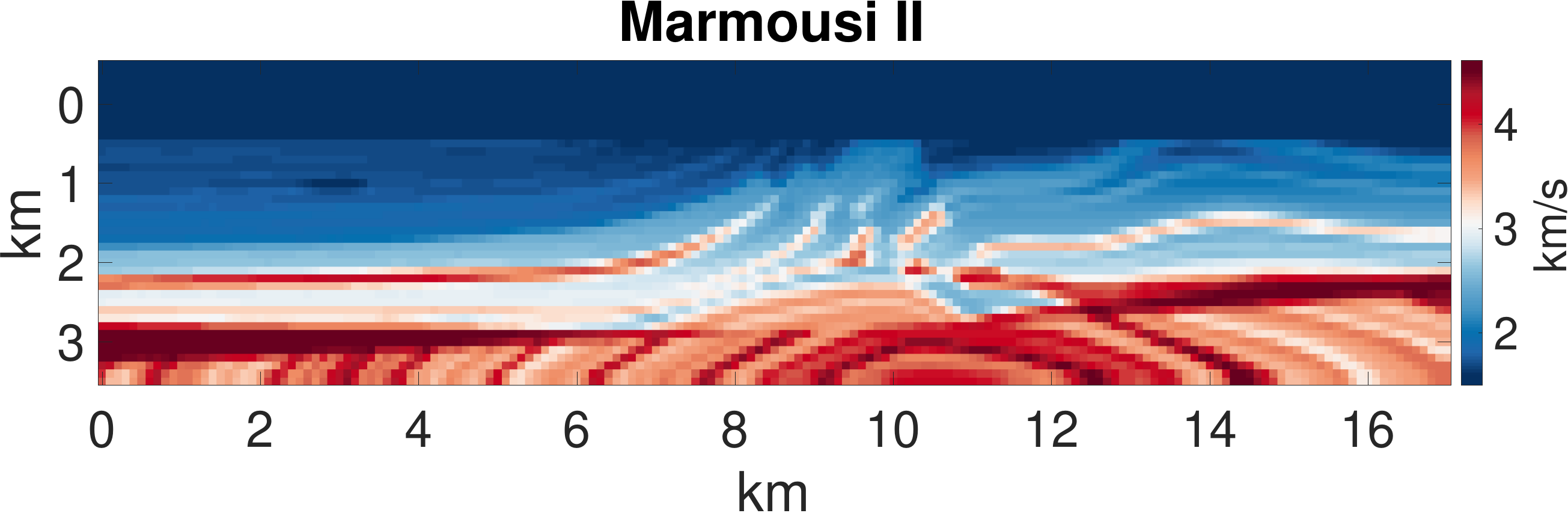}
    \includegraphics[width=0.44\linewidth]{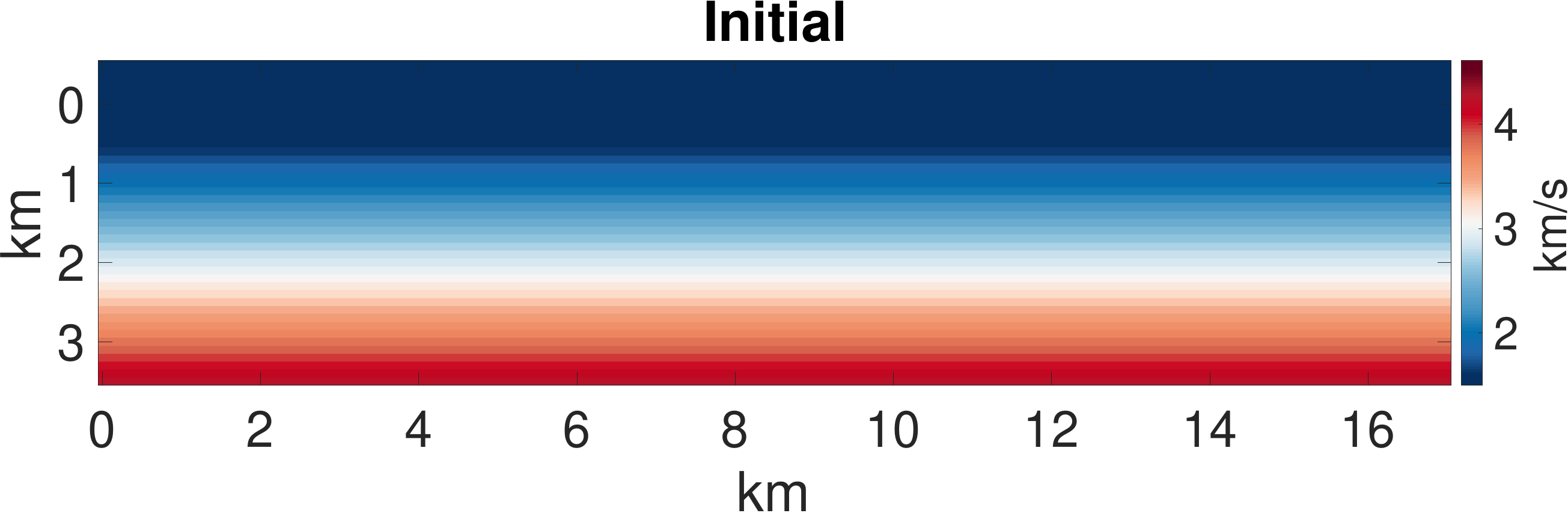}
    \includegraphics[width=0.44\linewidth]{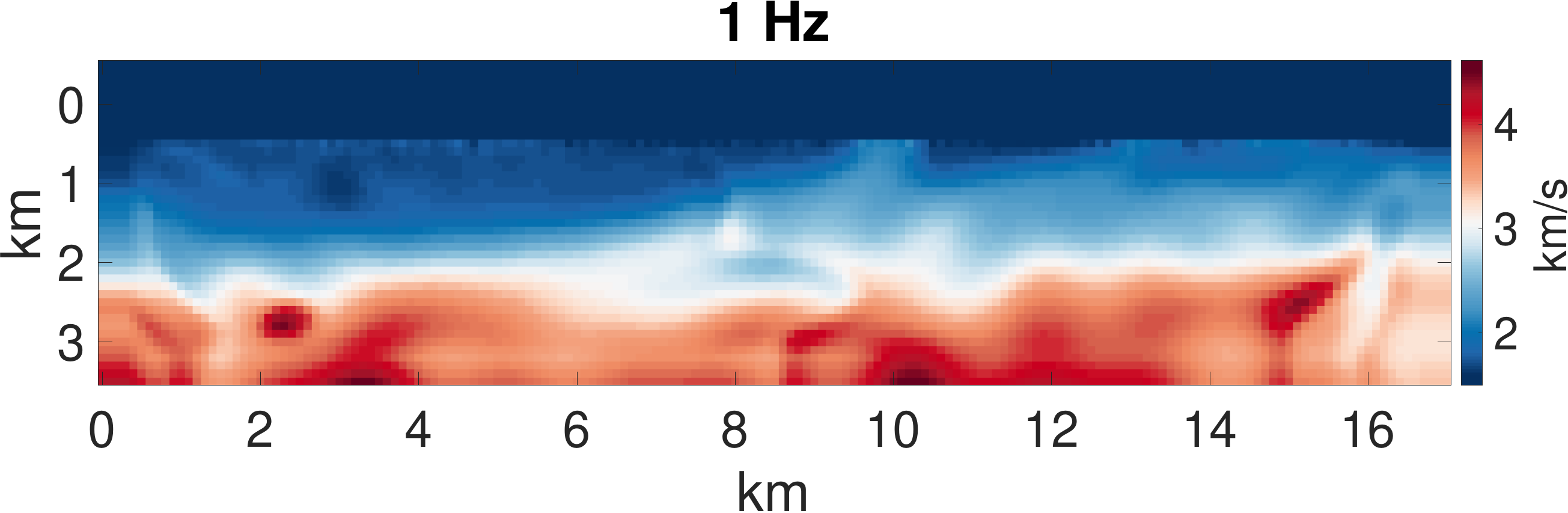}
    \includegraphics[width=0.44\linewidth]{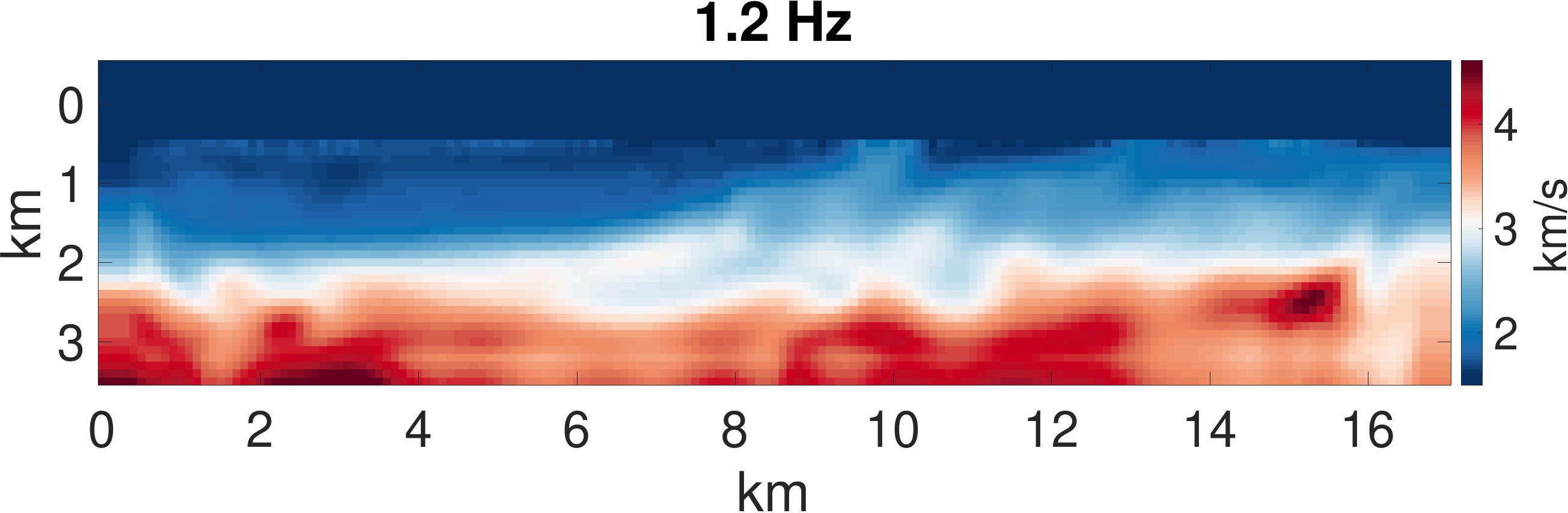}
    \includegraphics[width=0.44\linewidth]{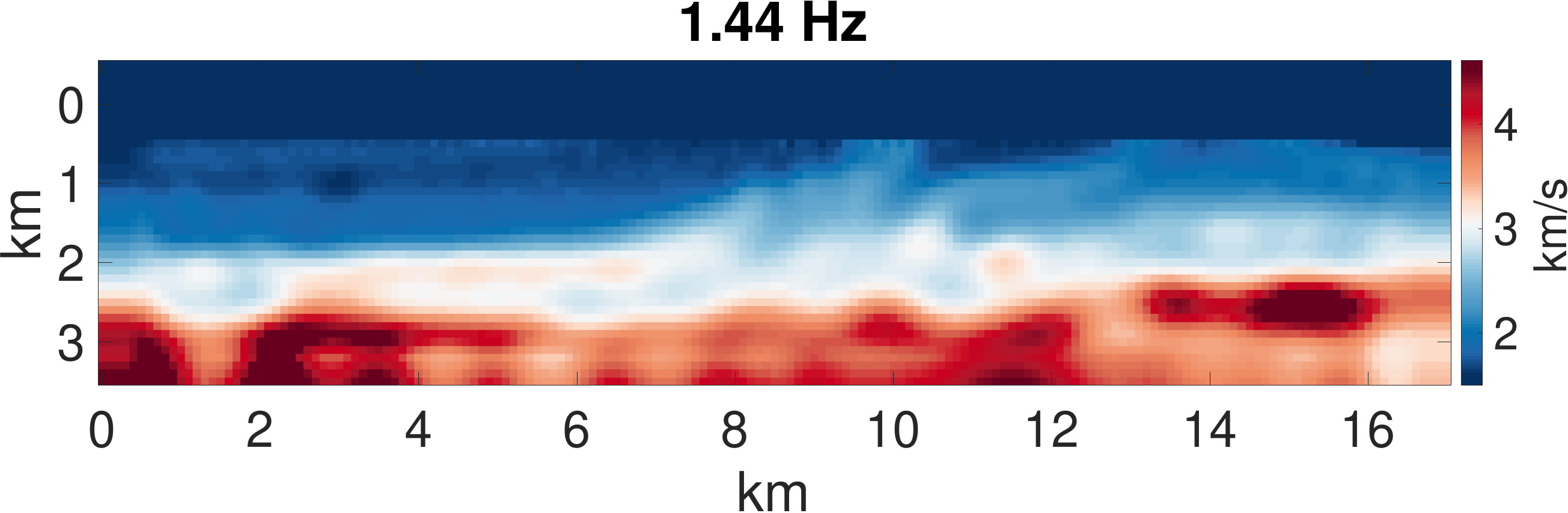}
    \includegraphics[width=0.44\linewidth]{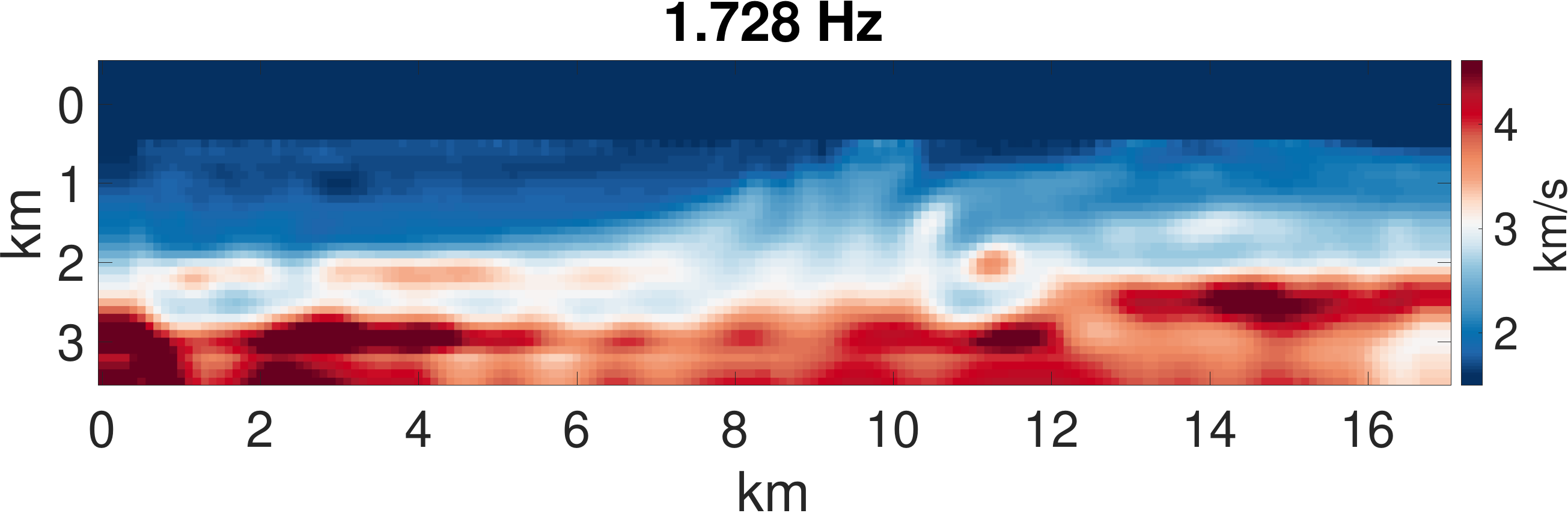}
    \includegraphics[width=0.44\linewidth]{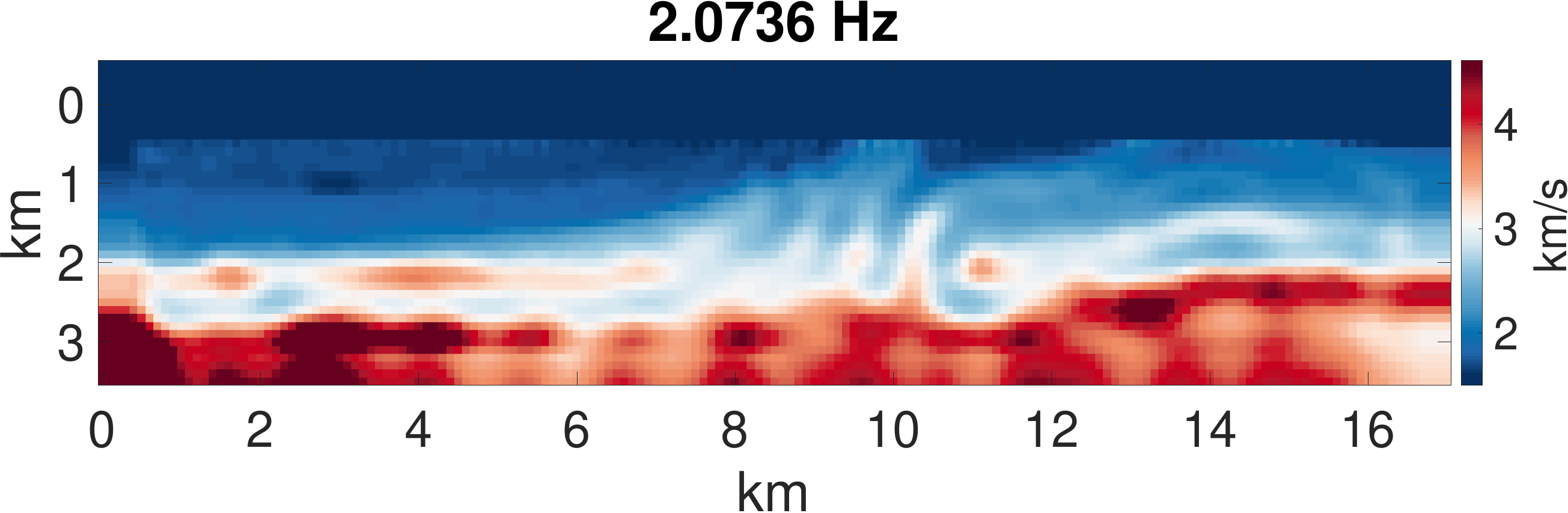}
    \includegraphics[width=0.44\linewidth]{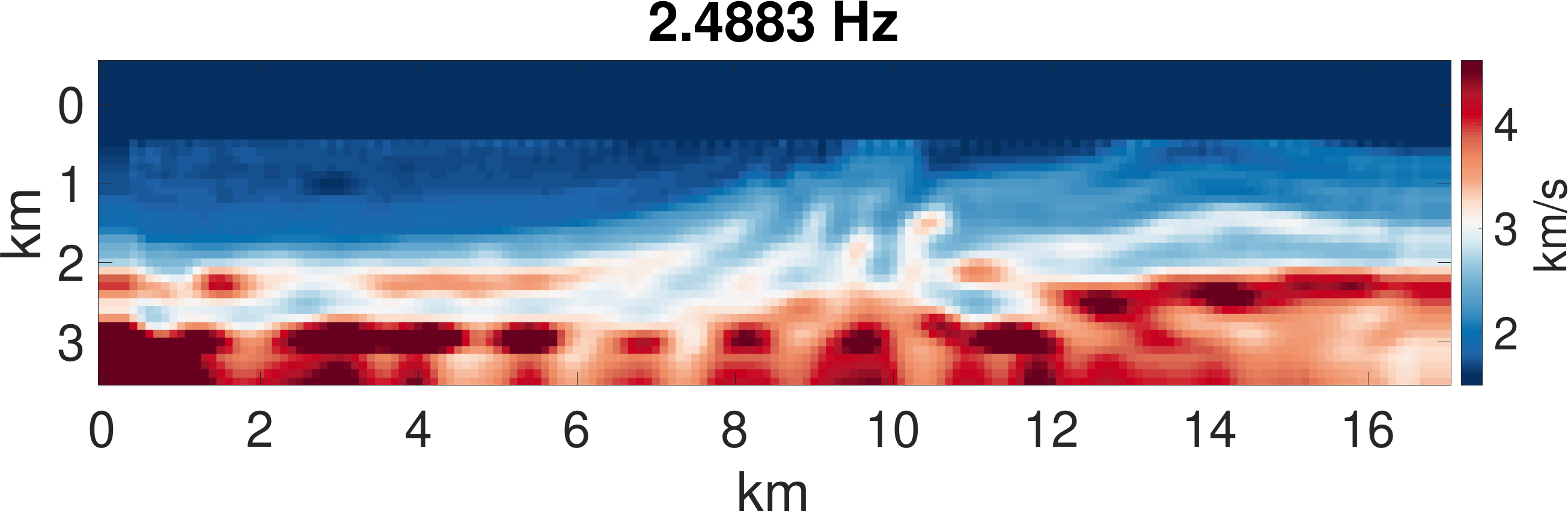}
    \includegraphics[width=0.44\linewidth]{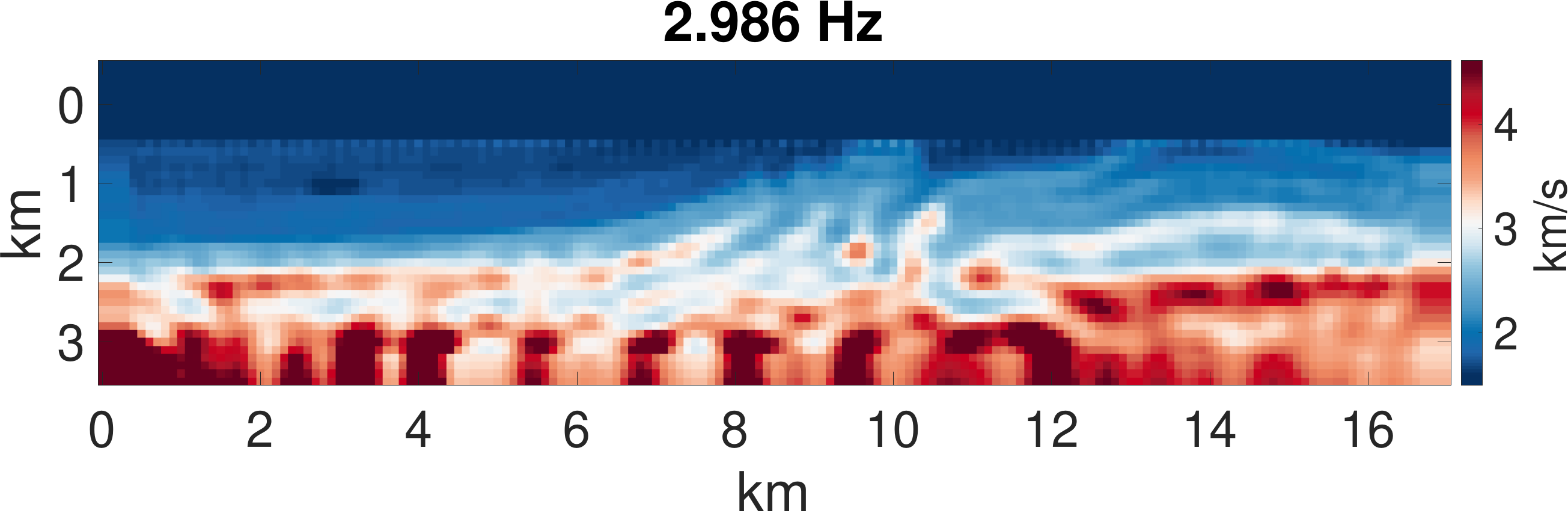}
    \includegraphics[width=0.44\linewidth]{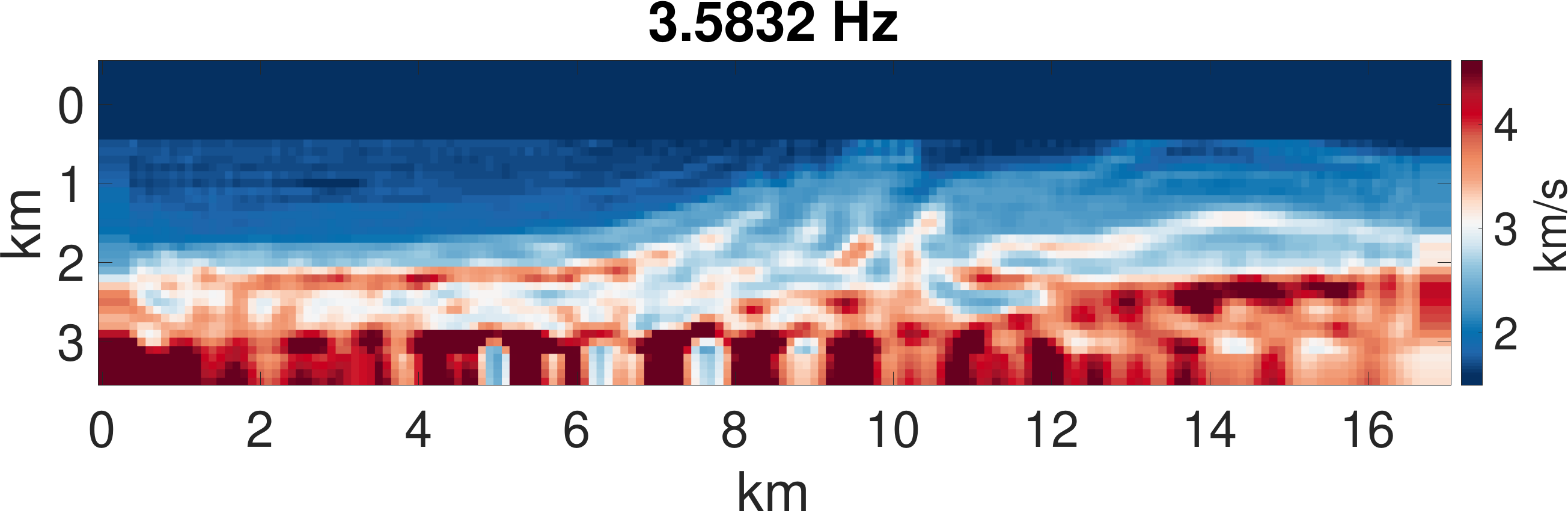}
    \includegraphics[width=0.44\linewidth]{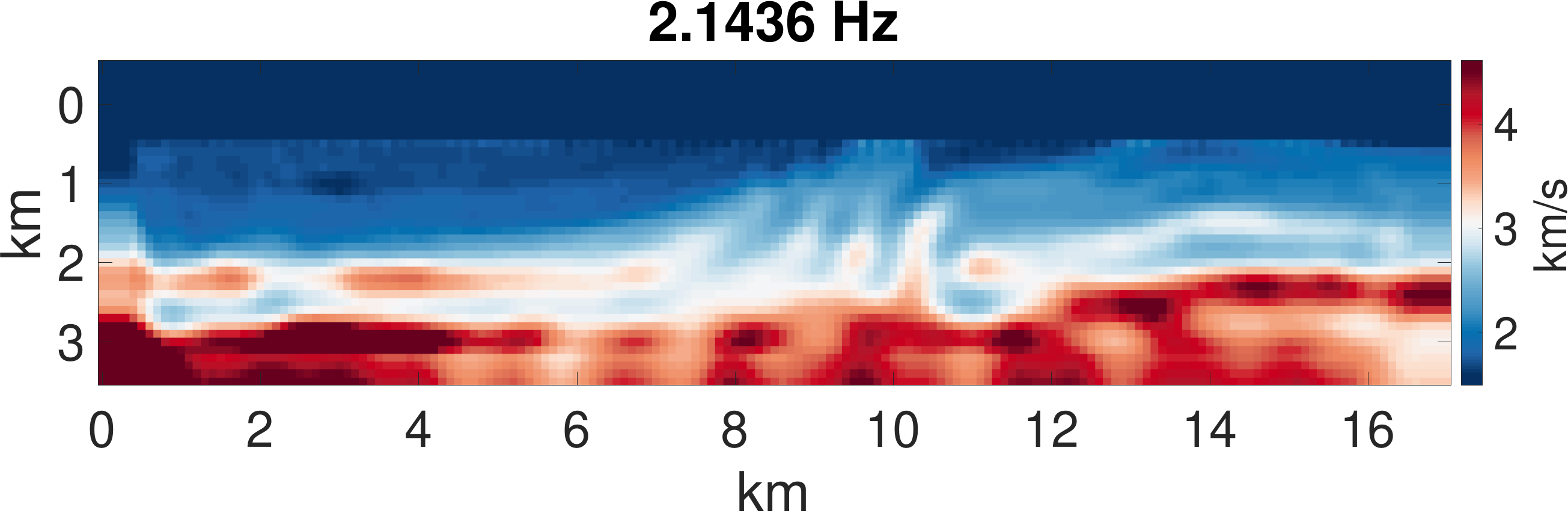}
    \includegraphics[width=0.44\linewidth]{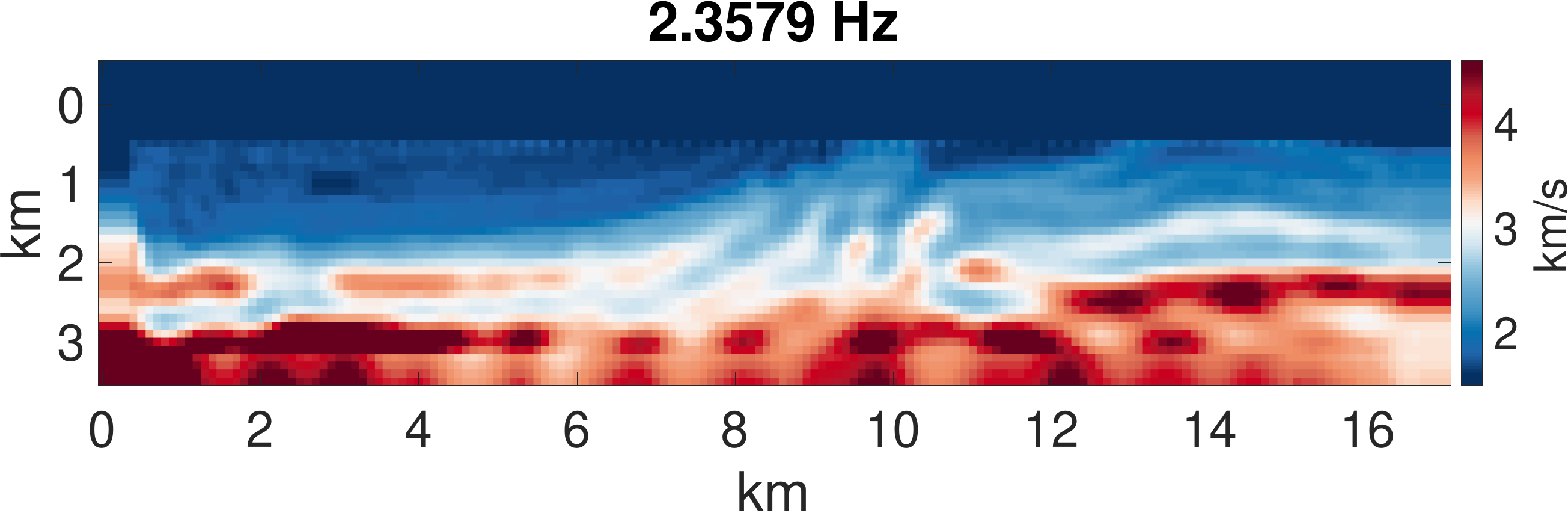}
    \includegraphics[width=0.44\linewidth]{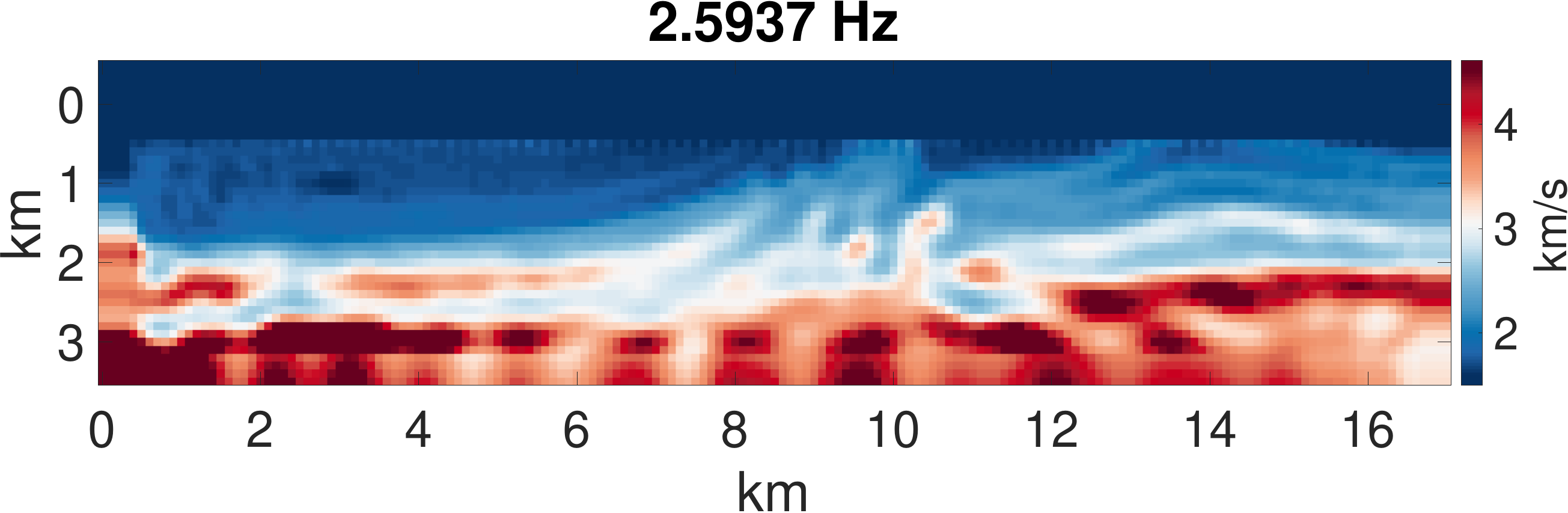}
    \includegraphics[width=0.44\linewidth]{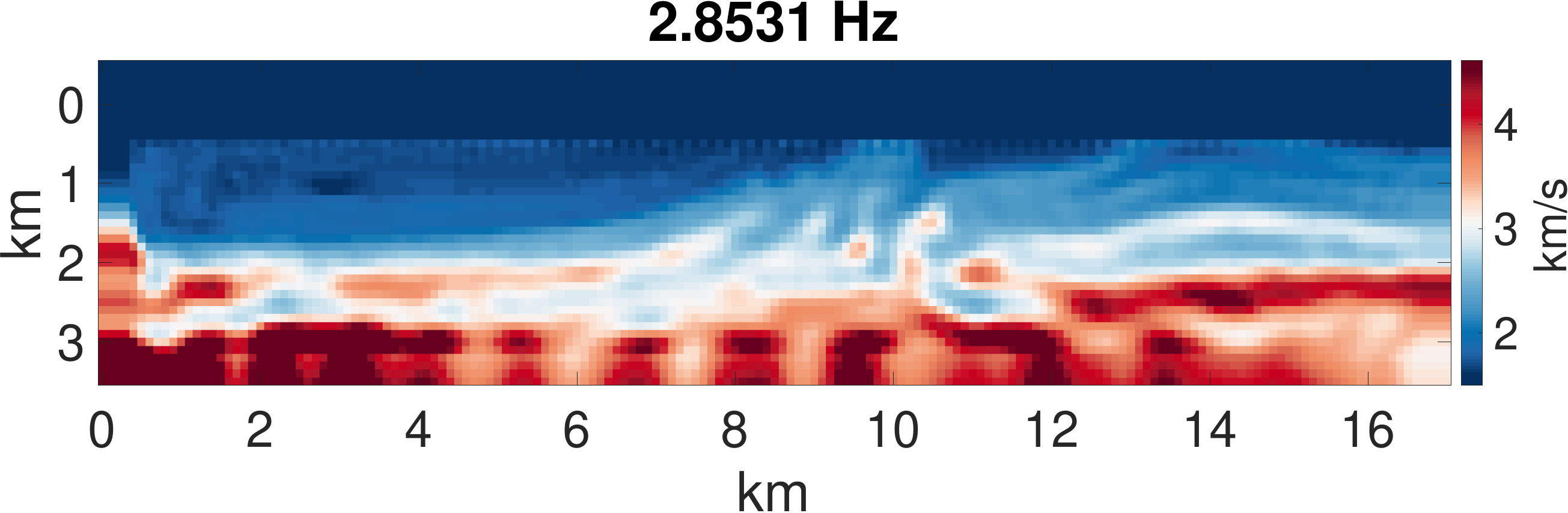}
    \includegraphics[width=0.44\linewidth]{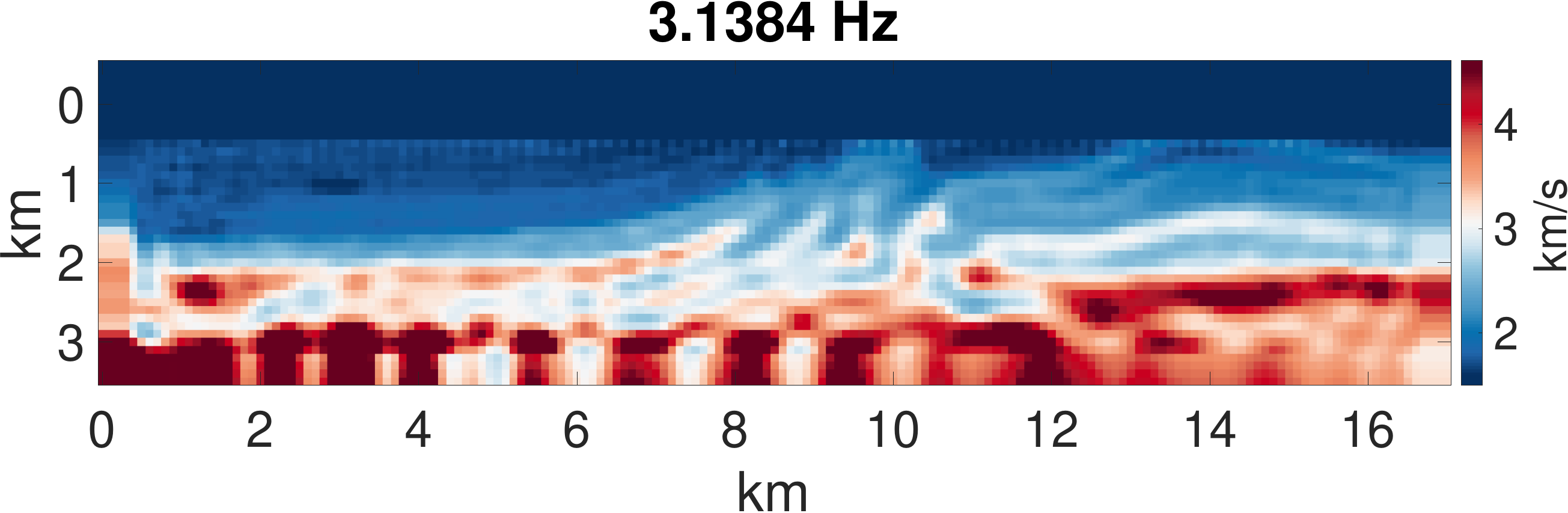}
    \includegraphics[width=0.44\linewidth]{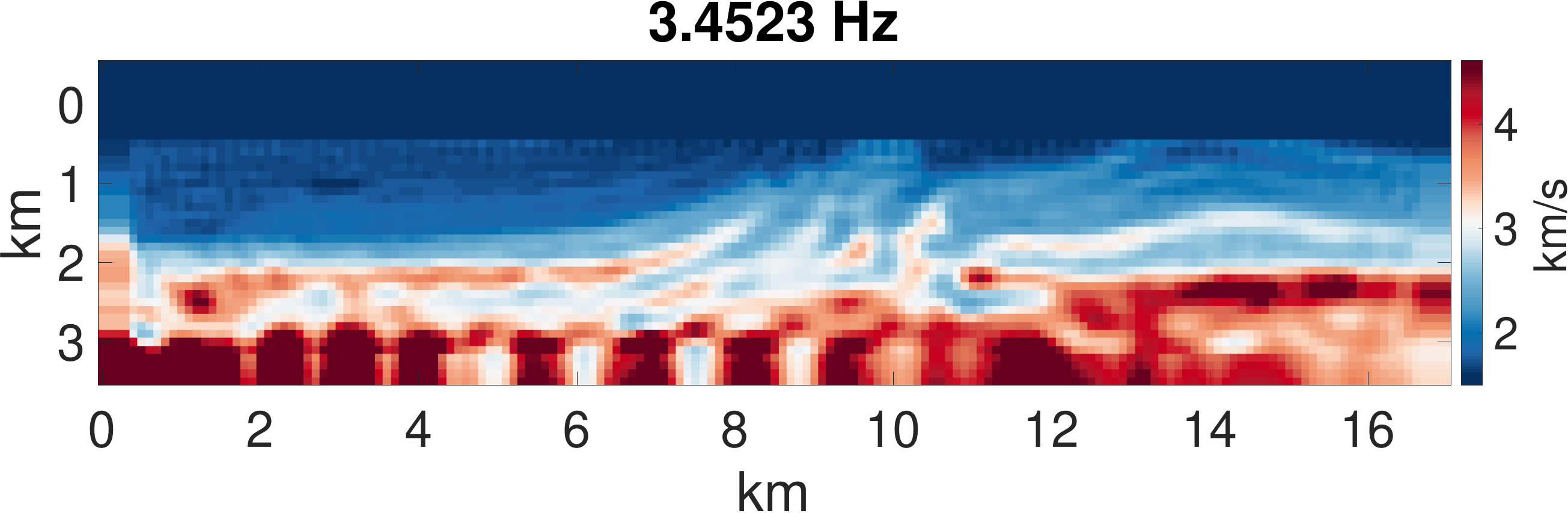}
    \includegraphics[width=0.44\linewidth]{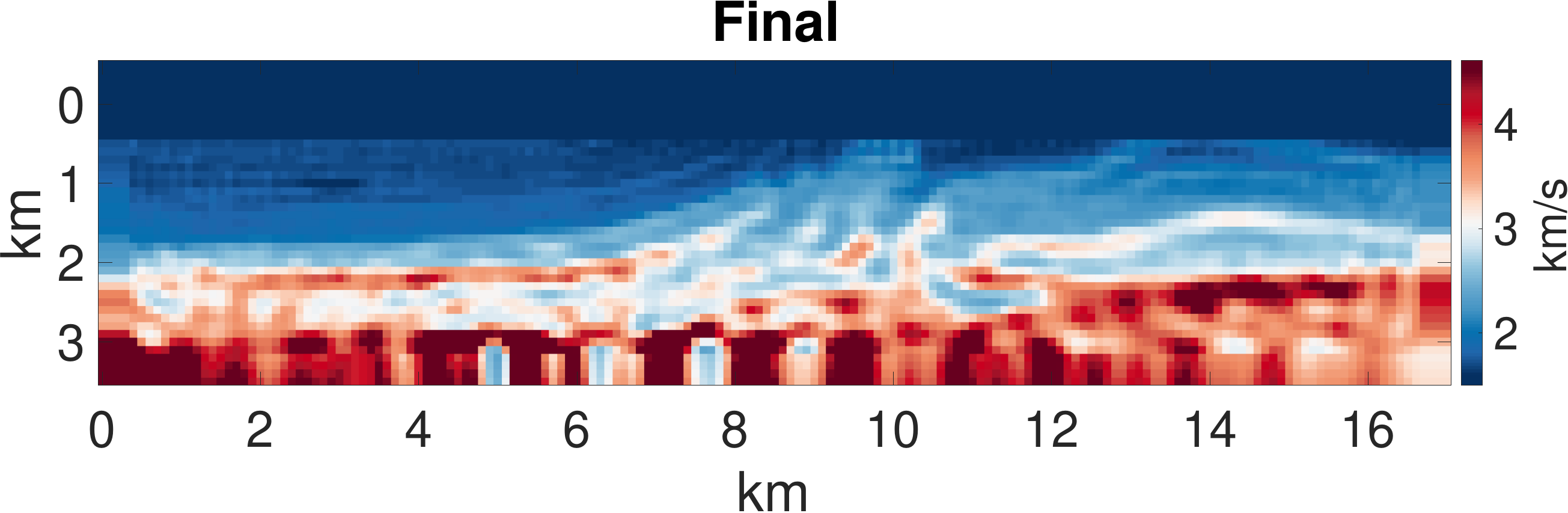}
    \caption[Classical FWI frequency updates.]{Classical FWI frequency updates. Starting from 1\si{Hz}, model update frequency was increased by a factor of 1.2 until a maximum frequency of 3.45\si{Hz}.The optimization algorithm was L-BFGS-B, with 50 iterations per step.}
    \label{fig:classical_fwi_progression}
\end{figure}

\subsection{Ray-Tracing}
Pre-cursor to FWI is ray-tracing modelling to assess areas of update from standard FWI formulation. Open source version of \textbf{fteikpy} Python library provided by \cite{Noble2014} was adapted and utilized on the Marmousi-2. This implementation computes accurate first arrival travel-times in 2D heterogeneous isotropic velocity models. The algorithm solves a hybrid Eikonal formulation with a spherical approximation near-source and a plane wave approximation in the far field. This reproduces properly the spherical behaviour of wave fronts in the vicinity of the source \cite{Noble2014}. Figure~\ref{fig:fteikpy_ray_tracing_marmousi} shows a sample of ray-paths for a source at 0km and depth 0km and ray coverage for the Marmousi model. 
\begin{figure}[ht!]
        \centering
        \subfloat[Sample of ray-paths through Marmousi]{\includegraphics[width=0.8\linewidth]{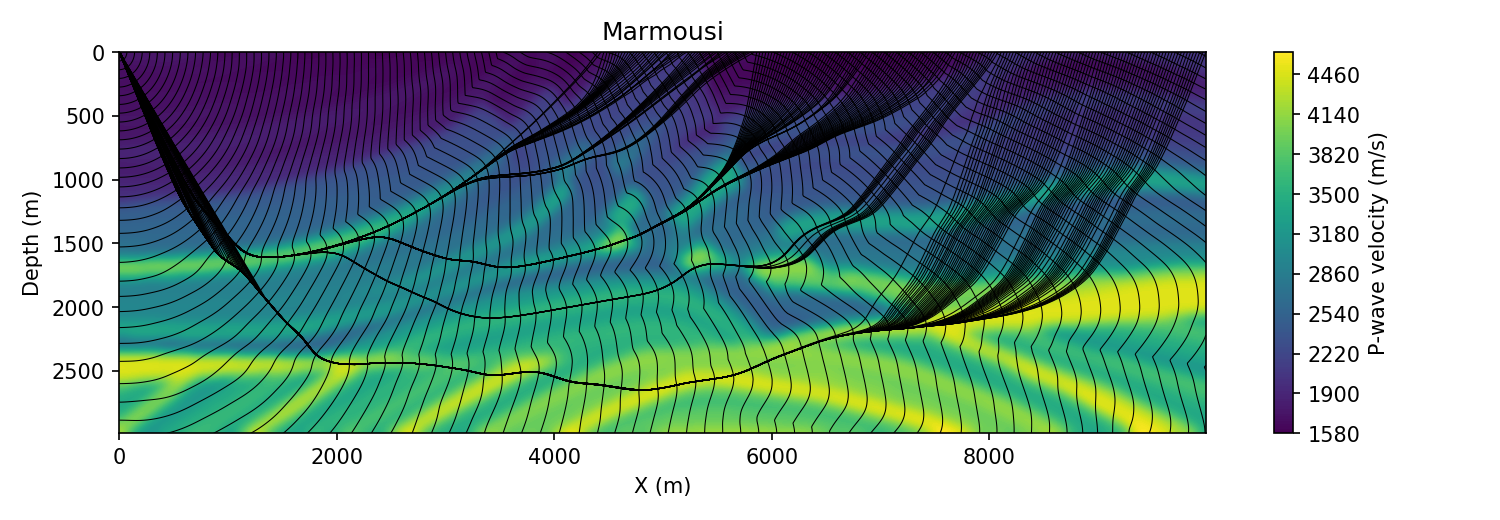}}
        \\
        \subfloat[Area of coverage intensity from ray-tracing.]{\includegraphics[width=0.8\linewidth]{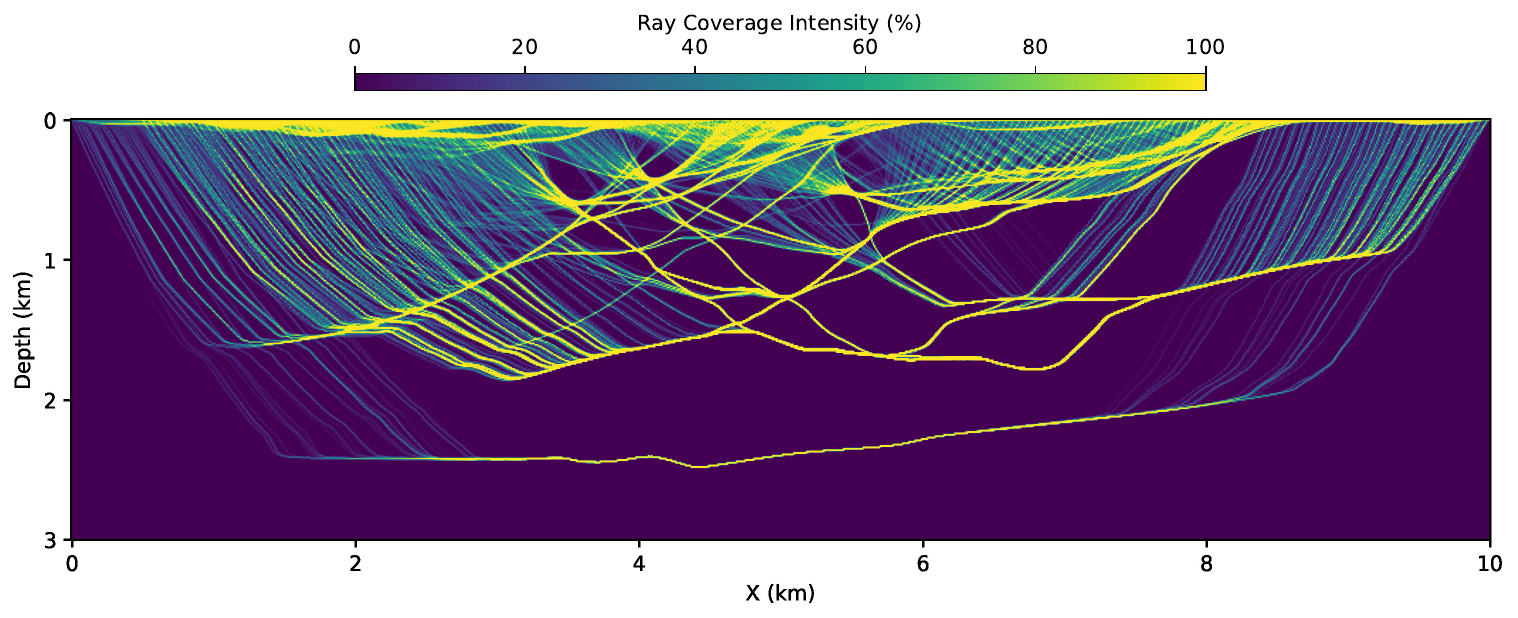}}
        \caption[Ray-tracing using \textbf{fteikpy}.]{Ray-tracing using \textbf{fteikpy}.}
        \label{fig:fteikpy_ray_tracing_marmousi}
\end{figure}

\clearpage
\bibliographystyle{unsrt}  
\bibliography{references}  %%% Remove comment to use the external 

\end{document}